\documentclass[apj]{emulateapj}
\usepackage{apjfonts}
\usepackage{epstopdf}
\usepackage{amsmath}

\def\wig#1{\mathrel{\hbox{\hbox to 0pt{%
          \lower.5ex\hbox{$\sim$}\hss}\raise.4ex\hbox{$#1$}}}}

\shorttitle{Water Clouds}

\newcommand{\teff}{$T_{\rm eff}$}

\newcommand{\cp}{\citep}
\newcommand{\ct}{\citet}

\newcommand{\icarus}{Icarus} 
\newcommand{\fsed}{$f_{\rm sed}$} 
\newcommand{\nas}{Na$_2$S}

\newcommand{\kzz}{$K_{zz}$}
\newcommand{\jwst}{\emph{JWST}}
\slugcomment{Draft for ApJ}

\begin{document}

\title{Water Clouds in Y Dwarfs and Exoplanets}

\author{Caroline V. Morley\altaffilmark{1,2}, Mark S. Marley\altaffilmark{3},  Jonathan J. Fortney\altaffilmark{1}, Roxana Lupu\altaffilmark{3}, Didier Saumon\altaffilmark{4}, Tom Greene\altaffilmark{3}, Katharina Lodders\altaffilmark{5}}

\altaffiltext{1}{Department of Astronomy and Astrophysics, University of California, Santa Cruz, CA 95064; cmorley@ucolick.org}
\altaffiltext{2}{Harriet P. Jenkins Graduate Student Fellow} 
\altaffiltext{3}{NASA Ames Research Center} 
\altaffiltext{4}{Los Alamos National Lab} 
\altaffiltext{5}{Washington University in St Louis} 

\begin{abstract}
The formation of clouds affects brown dwarf and planetary atmospheres of nearly all effective temperatures. Iron and silicate condense in L dwarf atmospheres and dissipate at the L/T transition. Minor species such as sulfides and salts condense in mid-late T dwarfs. For brown dwarfs below \teff$\sim$450 K, water condenses in the upper atmosphere to form ice clouds. Currently over a dozen objects in this temperature range have been discovered, and few previous theoretical studies have addressed the effect of water clouds on brown dwarf or exoplanetary spectra. Here we present a new grid of models that include the effect of water cloud opacity. We find that they become optically thick in objects below \teff$\sim$350--375 K. Unlike refractory cloud materials, water ice particles are significantly non-gray absorbers; they predominantly scatter at optical wavelengths through \emph{J} band and absorb in the infrared with prominent features, the strongest of which is at 2.8 \micron. H$_2$O, NH$_3$, CH$_4$, and H$_2$ CIA are dominant opacity sources; less abundant species such as may also be detectable, including the alkalis, H$_2$S, and PH$_3$. PH$_3$, which has been detected in Jupiter, is expected to have a strong signature in the mid-infrared at 4.3 \micron\ in Y dwarfs around \teff=450 K; if disequilibrium chemistry increases the abundance of PH$_3$, it may be detectable over a wider effective temperature range than models predict. We show results incorporating disequilibrium nitrogen and carbon chemistry and predict signatures of low gravity in planetary-mass objects. Lastly, we make predictions for the observability of Y dwarfs and planets with existing and future instruments including the \emph{James Webb Space Telescope} and \emph{Gemini Planet Imager}.

\end{abstract}

 
\section{Introduction}

Brown dwarfs link planetary and stellar astrophysics, with compositions like stars but the temperatures of planets. They form the tail of the initial mass function and, too low in mass to have core temperatures high enough to fuse hydrogen, they cool over time through the brown dwarf spectral sequence. As they cool, different molecules and condensates form and carve their spectra. 

With the discovery of very cool brown dwarfs we are able to investigate for the first time the physical and chemical processes that occur in atmospheres with effective temperature ranges that would be suitable for a warm beverage. While brown dwarfs are free-floating, they should share many of the same physical processes as the giant planets that will be uncovered by future surveys. 

\subsection {Discovery and Characterization of Y Dwarfs}

The proposed spectral class Y encompasses brown dwarfs that have cooled below \teff$\sim$500 K. About 17 objects have been classified as Y dwarfs to date. Many of those have now been found using the Wide-field Infrared Survey Explorer (WISE) \cp{Cushing11, Kirkpatrick12, Liu12, Tinney13, Kirkpatrick13}. Additional objects have been discovered as wide-separation companions. \ct{Liu11} found a very cool ($\sim$Y0) companion to a late T dwarf; \ct{Luhman12} discovered a $\sim$300--350 K object orbiting a white dwarf. At these temperatures, NH$_3$ absorption features begin to appear in their near-infrared spectra, and sodium and potassium wane in importance in the optical as they condense into clouds. 

Recent follow-up studies have aimed to characterize the Y dwarf population. There has been a large effort to measure parallaxes of Y dwarfs: \ct{Marsh13} present results for 5 Y dwarfs and 3 late T dwarfs using a compilation of data from the ground and space. \ct{Dupuy13} present results using only the \emph{Spitzer} Space Telescope for 16 Y and T dwarfs. \ct{Beichman14} present results from a compilation of data from Keck II, the \emph{Spitzer} Space Telescope, and the \emph{Hubble} Space telescope for 15 Y and T dwarfs. Groups have also been collecting followup observations to better understand the spectral energy distributions of Y dwarfs. \ct{Leggett13} present followup near-infrared photometry for six Y dwarfs and a far-red spectrum for  WISEPC J205628.90+145953.3. \ct{Lodieu13} observed 7 Y dwarfs in the \emph{z} band using the Gran Telescopio de Canarias. 

\subsection{Previous Models of Y dwarfs}

 A number of available models for brown dwarfs include models cold enough to represent Y dwarfs \cp{Allard12, Saumon12, Morley12}, but do not yet treat the effects of water clouds. The first models to incorporate the effects of water clouds into a brown dwarf atmosphere are those of \ct{Burrows03b}. These models generally find that water clouds do not strongly affect the spectrum of Y dwarfs, but there have been few followup studies. \ct{Hubeny07} also includes simple water clouds with a fixed mode particle size of 100 \micron. In Section \ref{burrows} we will discuss how our results compare to these early models. 

A number of studies have included water clouds in exoplanetary atmospheres; \ct{Marley99} and \ct{Sudarsky00} both modeled the effect of water clouds on the albedos of giant exoplanets; they find the formation of water clouds significantly increases the planetary albedos. \ct{Burrows04} also consider water clouds in exoplanets, using a similar approach as \ct{Burrows03b} but for irradiated planets; \ct{Sudarsky03} and \ct{Sudarsky05} calculate the thermal emission of exoplanets that include water clouds and find that they have a strong effect on the emergent spectrum.

\subsection{Clouds in L and T dwarfs}

Clouds have posed the greatest challenge for brown dwarf modeling since the first L dwarfs were discovered. As brown dwarfs cool along the L sequence from 2500 K to 1300 K, refractory materials like corundum, iron, and silicates condense to form thick dust layers \cp{Lunine86, Fegley96, Burrows99, Lodders02, Lodders03, Lodders06, Helling06, Visscher10} which thicken as the brown dwarf cools. These dust clouds shape the emergent spectra of L dwarfs \cp[see, e.g.][]{Tsuji96, Allard01,Marley02,Burrows06,Helling08,Cushing08,Witte11}. 

For field brown dwarfs these clouds clear over a very small range of effective temperature around 1200--1300 K and around the same temperature, methane features begin to appear in the near-infrared. The brown dwarf is then classified as a T dwarf and, for many years, mid to late T dwarfs were considered to be cloud-free. However, it has long been recognized that other somewhat less refractory materials such as sulfides and salts should condense in cooler T dwarfs \cp{Lodders99}. As late T dwarfs (500--900 K) were discovered and characterized, a population of objects redder in the near-infrared (e.g. $J-K$, $J-H$ colors) than the predictions of cloud-free models emerged. \ct{Morley12} included the clouds predicted to form by condensation of the sulfides and alkali salts and showed that by including thin layers of these clouds, the colors and spectra of these redder observed T dwarfs can be matched. As these T dwarfs are further characterized, variability has been observed in mid-late T dwarfs \cp{Buenzli12}; the sulfide clouds may play a role in this variability. 

\subsection{Directly-imaged Exoplanets}

Spectra of directly-imaged planets are also strongly influenced by the opacity of clouds. The first multi-planet directly-imaged system, HR 8799, has four planets, all of which have infrared colors that indicate cloudy atmospheres, much like L dwarfs \cp{Marois08}. Other planetary-mass objects also appear to have spectral properties similar to L dwarfs including {$\beta$} Pictoris b \cp{Bonnefoy13} and 2M1207b \cp{Barman11b}. In fact, at similar effective temperatures planetary-mass objects appear to be even more cloudy than their brown dwarf counterparts \cp{Madhu11c, Barman11}, which has been used to suggest that the breakup of the iron and silicate clouds at the L/T transition may be gravity dependent \cp{Metchev06, Marley12}. 

Nonetheless the transition to a methane-dominated atmosphere and cloud-depleted near-infrared spectrum must happen at some effective temperature, with the resulting objects appearing as low-gravity `T' and `Y' dwarfs. The first such object discovered is GJ 504b \cp{Kuzuhara13} which is currently the coldest directly-imaged planet (\teff$\sim$500 K) and has colors very similar to T dwarfs; followup observations probing the methane feature at 1.6 \micron\ suggest that, as expected from thermochemical equilibrium calculations, methane is present in the atmosphere \cp{Janson13}. 

\subsection{Water clouds}
In a cold solar composition atmosphere, water clouds will be a massive cloud and an important opacity source. Unlike the refractory clouds which have been extensively studied by a number of groups \cp{AM01, Helling06, Allard01, Tsuji96, Burrows06, Helling08}, the same attention has not been paid to volatile clouds in brown dwarfs. In this work, we aim to predict the effects that water clouds will have on brown dwarf atmospheres. We calculate pressure--temperature profiles, spectra, and colors for the coolest brown dwarfs. We study the signatures of the water clouds and their optical properties, estimate their likely particle sizes, and determine at which effective temperatures the cloud will become optically thick in the photosphere. We end by considering the observability of Y dwarfs with the four major instruments being built for the \emph{James Webb Space Telescope (JWST)} and the detectability of cool giant planets with new and upcoming ground-based instruments. 
 
\section{Methods}

\subsection{Atmosphere Model}

We calculate 1D pressure--temperature profiles, which are self-consistent with both the chemistry and clouds, of atmospheres in radiative--convective equilibrium. The thermal radiative transfer is determined using the ``source function technique'' presented in \ct{Toon89}. The gas opacity is calculated using correlated-k coefficients to increase calculation speed; our opacity database incorporates published data from both laboratory experiments and first-principles quantum mechanics calculations and is described extensively in \ct{Freedman08}. The opacity database includes two significant updates since \ct{Freedman08}, which are described in \ct{Saumon12}: a new molecular line list for ammonia \cp{Yurchenko11} and an improved treatment of the pressure-induced opacity of H$_2$ collisions \cp{Richard12}. The cloud opacity is included as Mie scattering of spherical cloud particles in each atmospheric layer. The atmosphere models are more extensively described in \ct{Mckay89, Marley96, Burrows97, MM99, Marley02, Saumon08, Fortney08b}.

\subsection{Cloud Model}

The cloud model calculates the vertical locations, heights, and mode particle sizes of clouds as they condense in the atmosphere. The calculation is coupled with the radiative transfer calculations, so a converged model will have a temperature structure that is self-consistent with the clouds. 

The cloud code is a modification of the \ct{AM01} cloud model. This model has successfully been used to model the effects of the iron, silicate, and corundum clouds on the spectra of L dwarfs \cp{Saumon08,Stephens09} as well as the sulfide and chloride clouds that likely form in the atmospheres of T dwarfs \cp{Morley12}. Here, we modify it to include the effects of water clouds. The \ct{AM01} approach avoids treating the highly uncertain microphysical processes that create clouds in brown dwarf and planetary atmospheres. Instead, it aims to balance the advection and diffusion of each species' vapor and condensate at each layer of the atmosphere. It balances the upward transport of vapor and condensate by turbulent mixing in the atmosphere with the downward transport of condensate by sedimentation. This balance is achieved using the equation 

\begin{equation} \label{advdiff}
-K_{zz} \frac{\partial q_t}{ \partial z} - f_{\textrm{sed}} w_*q_c =0, 
\end{equation}
where $K_{zz}$ is the vertical eddy diffusion coefficient, $q_t$ is the mixing ratio of condensate and vapor, $q_c$ is the mixing ratio of condensate, $w_*$ is the convective velocity scale, and \fsed\ is a parameter that describes the efficiency of sedimentation in the atmosphere. 

Solving this equation allows us to calculate the total amount of condensate in each layer of the atmosphere. We calculate the modal particle size using the sedimentation flux and by prescribing a lognormal size distribution of particles, given by
\begin{equation}
\dfrac{dn}{dr}=\dfrac{N}{r\sqrt{2\pi}\ln\sigma}\exp{\left[\dfrac{\ln^2(r/r_g)} {2\ln^2\sigma}\right]}
\end{equation}
where $N$ is the total number concentration of particles, $r_g$ is the geometric mean radius, and $\sigma$ is the geometric standard deviation. We fix $\sigma$ at 2.0 for this study (see discussion in \ct{AM01}). We calculate the falling speeds of particles within this distribution assuming viscous flow around spheres (and using the Cunningham slip factor to account for gas kinetic effects). We calculate the other parameters in equation \ref{advdiff} ($K_{zz}$ and $w_*$) using mixing length theory and by prescribing a lower bound for $K_{zz}$ of 10$^5$ cm$^2$/s, which represents the residual turbulence due to processes such as breaking gravity waves in the radiative regions of the atmosphere. 

This process allows us to calculate the mode particle size in each layer of the atmosphere using calculated or physically motivated values for all parameters except for the free parameter \fsed. In general, we find larger particles (which have higher terminal velocities) in the bottom layers of a cloud and smaller particles (which have lower terminal velocities) in the upper layers.

A high sedimentation efficiency parameter \fsed\ results in vertically thinner clouds with larger particle sizes, whereas a lower \fsed\ results in more vertically extended clouds with smaller particles sizes. As a result, a higher \fsed\ corresponds to optically thinner clouds and a lower \fsed\ corresponds to optically thicker clouds. 

The \ct{AM01} cloud model code computes the available quantity of condensible gas above the cloud base by comparing the local gas abundance (accounting for upwards transport by mixing via $K_{\rm zz}$) to the local condensate vapor pressure $p_{\rm vap}$. In cases where the formation of condensates does not proceed by homogeneous condensation we nevertheless compute an equivalent vapor pressure curve.

\subsection{Challenges of water clouds}

The condensation of water vapor into water ice clouds poses some unique problems for our self-consistent equilibrium approach. Water vapor is the most dominant source of opacity in a Y dwarf atmosphere. After it condenses, if the cloud opacity is somehow removed, there is very little gas opacity left in the atmosphere and the layers beneath can radiate efficiently. However, the cloud opacity cannot just be removed from the atmosphere; it must condense into a cloud, and because oxygen is one of the most abundant elements, the water cloud that forms is quite massive and optically thick. This means that the dominant vapor opacity source condenses into a dominant solid opacity source. 

In practice, when aiming to calculate a solution in radiative--convective and chemical equilibrium that is self-consistent with the water cloud, we find that we are not able to find a self-consistent solution for a range of model effective temperatures from $\sim$225--450 K. As the water cloud forms in the model, it significantly warms the atmosphere below it because it prevents flux from escaping. This warming causes the cloud to evaporate, removing the opacity source, and allowing flux to escape and cool the atmosphere again; a cloud forms again. An equilibrium solution is not found. 

To solve this problem, we borrow phenomenological ideas from our own solar system. When water clouds are observed in the solar system on Earth, Jupiter, and Saturn, they never form in a globally homogeneous layer. Instead, they form into patchy clouds. For example, on Jupiter, there are 5 \micron\ hot spots though which flux emerges from deep within the atmosphere; these are believed to be ``holes'' or thin areas of the deep Jovian water cloud \cp{Westphal69, Westphal74, Orton96, Carlson94}. Saturn has similar mid-infrared heterogeneity \cp{Baines05}. 

Because water clouds \emph{never} appear to form globally homogeneous, uniform clouds in the solar system planets, models that include clouds for these planets do not attempt to find a 1D steady-state equilibrium solution in a self-consistent way. Instead, for clouds in Earth's atmosphere, the evolution of clouds is either modeled over time using a time-stepping model or the clouds are modeled in 2D or 3D. In fact a method sometimes used in 3D circulation models on Earth inspires our approach, described below. In these circulation models, clouds form on scales smaller than the grid scale; cloud opacity is implemented using a two-column sub-grid approach. Other previous efforts for exoplanets and brown dwarfs either did not iterate or used a highly specified cloud parametrization for particle size and cloud height \cp[e.g][]{Marley99, Sudarsky00, Hubeny07}. Our approach here is not to include a specified cloud, but instead to iteratively solve for a cloud profile that is self-consistent with the atmosphere structure. We rely on a model which successfully reproduces cloud particle sizes and distributions on Jupiter and Earth \cp{AM01}. Perhaps fortuitously, the particular cloud model employed by \ct{Burrows03b}, who also solve the problem self-consistently, produced somewhat large particles with small infrared optical depth. More general cases in which some clouds have smaller particle sizes have a far greater optical depth and are more challenging to converge.

In this work, we make the assumption that patchy water clouds also form in brown dwarfs. Theoretical motivation for this patchiness has not yet been well-developed, but recent highly idealized models suggest that the rotation and internal heating of brown dwarfs could drive jet-like or vortex-like circulation \cp{Zhang14}. Such weather patterns in the atmosphere may create rising and sinking parcels of air and maintain inhomogeneous clouds. 

Other evidence for cloud patchiness in the silicate cloud decks of warmer brown dwarfs has been seen in the variability observations by, e.g., \ct{Radigan12} and the spatial mapping of the L/T transition brown dwarf Luhman 16B \cp{Crossfield14}.

In this model, even though the water cloud layer forms a thick opacity source, the flux can emerge from holes in the cloud deck. This assumption allows us to calculate a temperature structure in radiative--convective equilibrium that is self-consistent with the clouds because flux is able to emerge through the holes even when the clouds become optically thick. This means that the temperature structure can remain cool enough to have a condensed water cloud layer. 

\subsection{Implementing patchy clouds} \label{partlycloudymodels}

 \begin{figure}[t]
  \center   \includegraphics[width=3.75in]{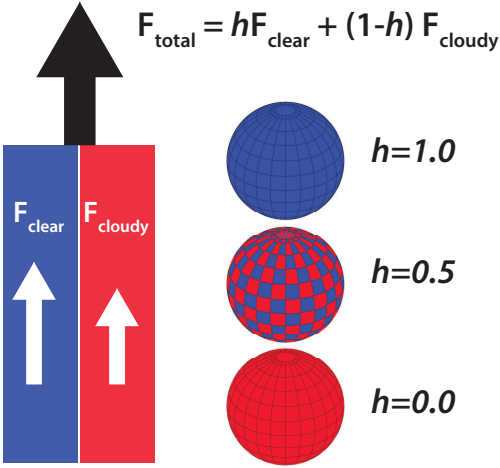}
 \caption{Partly cloudy model atmospheres. This cartoon illustrates our approach to calculating pressure--temperature profiles in radiative--convective equilibrium for partly cloudy atmospheres. We calculate the flux separately through two columns: one that does not include cloud opacity and one that does. We then sum these fluxes to calculate the total flux, according to the fraction of the surface we assume to be covered by holes, $h$. $h=1$ represents a fully clear atmosphere; $h=0$ represents a fully cloudy atmosphere. For the models in the grid presented here, $h=0.5$. }
\label{cartoon}
\end{figure}

We calculate patchy clouds following the approach of \ct{Marley10}, who implemented patchy clouds in an attempt to understand a mechanism that could reproduce the L/T transition, in which clouds break up progressively and more flux emerges from holes in the clouds. Following this prescription, we calculate flux separately through both a cloudy column (with the cloud opacity included) and a clear column (with the cloud opacity not included) with the same pressure--temperature profile. This calculation is shown schematically in Figure \ref{cartoon}. We can change the cloud-covering fraction by varying $h$, the fractional area of the atmosphere assumed to be covered in holes: 

\begin{equation}
F_{\rm total} = h F_{\rm clear} + (1-h) F_{\rm cloudy}
\end{equation}

Using this summed flux $F_{\rm total}$ through each atmospheric layer, we iterate as usual until we find a solution in radiative--convective equilibrium. Using $h\gtrsim$0.4--0.5 greatly improves convergence. With a hole fraction of at least 40--50\%, flux escapes from warm, deep layers through the cloud-free column and a consistent $P-T$ profile with water clouds can be calculated. 

\subsection{Cloud properties}

To model clouds and their radiative effect in the atmosphere, we need three pieces of information about the material. The first is the optical properties---the real and imaginary parts of the refractive index---of the condensed solid or liquid. In Y dwarfs, water always condenses in the solid phase \cp{Burrows03b}, so we use the optical properties of water ice \cp{Warren84}. The second property is the material's density; we use 0.93 g/cm$^3$ for water ice. Lastly, we need the saturation vapor pressure of water ice, which tells us where the cloud will form and how much material is available to form it. We assume that all material in excess of the saturation vapor pressure condenses to form a cloud. We use the equation from \ct{Buck81} to describe the saturation vapor pressure of water ice: 

\begin{equation}
 p_{\rm vap}= a  \exp\left[ \frac{(b - T_c/d)T_c} {T_c + c} \right] 
\end{equation}
where $T_c$ is the temperature in degrees Celsius and $a$, $b$, $c$, and $d$ are constants (6.1115$\times10^3$, 23.036, 279.82, and 333.7, respectively). 

The other clouds included in these models are Cr, MnS, \nas, KCl, and Zns. The thermochemical models that describe the formation of these clouds are described in \ct{Visscher06}. Using these models, fits to the saturation vapor pressure as a function of pressure, temperature, and metallicity were presented in \ct{Morley12}, Section 2.4. Sources of the optical properties used in the Mie scattering calculations are also presented in Table 1 of \ct{Morley12}. 

\subsection{Model grid} \label{modelgrid}

Our grid of models encompasses the full range of Y dwarfs and extends the grid from \ct{Morley12} to lower temperatures. This grid includes models from 200--450 K in increments of 25--50 K. We include surface gravities that range from giant planets to brown dwarfs, from log $g$ of 3.0 to 5.0 in increments of 0.5. We run all models on this main grid with \fsed=5 and $h$=0.5. 

Of course, it is likely that $h$ and \fsed\ vary, and could in fact be different for the underlying sulfide/salt clouds and the higher altitude water clouds. In fact, models at the L/T transition generally need to include non-uniformity in cloud properties across the transition to match its shape \cp{Marley10}. We do not aim to fully model all parts of this parameter space here. However, we do explore some parts of this space; we run additional models in which we vary \fsed\ from 3--7 at a single surface gravity (log $g$=4.0) and in which we vary $h$ from 0.2 to 1.0 (see Figure \ref{diffcloudcover}). We assume solar metallicity composition for all models, using elemental abundances from \ct{Lodders03}. 

The grid was deliberately chosen to be square, but includes some unphysical combinations of temperature and surface gravity, because higher mass brown dwarfs cannot have cooled enough during the age of the universe to reach very low temperatures. For the coolest models in our grid, \teff=200 K, the maximum expected log $g$ (that of a 10 Gyr brown dwarf) is $\sim4.2$. For 300 K, maximum log $g$ is $\sim4.5$--4.6; for 450 K, $\sim$4.7--5.0 \cp{Saumon08}. The ranges come from using a cloudy or cloud-free atmospheric boundary condition for the evolution models, see Figures 4 and 5 from \ct{Saumon08}. 

\subsection{Evolution models}

Absolute fluxes and magnitudes are calculated from our model spectra by applying the evolutionary radii of \ct{Saumon08}. Those cooling sequences provide the radius of the brown dwarf as a function of $T_{\rm eff}$ and $\log g$. Here we have used the radii from the evolution sequences computed with cloudless atmospheres as the surface boundary condition. A fully self-consistent calculation of the evolution would use a surface boundary condition defined by the corresponding model atmospheres. This is becoming increasingly difficult for brown dwarfs as the evolution of clouds during the long cooling time of the very cool objects considered here is rather complex. The sequence of transitions from cloudy L dwarfs to mainly clear mid-T dwarfs, to late-T dwarfs veiled with sulfide clouds (MnS, Na$_2$S, ZnS, Cr, KCl), which may also clear out in early Y dwarfs before water clouds appear, has yet to be understood properly, both empirically and theoretically. The use of a uniform, cloudless surface boundary condition has the virtue of simplicity. We can estimate the uncertainty in the radius thus obtained by comparing the surface boundary condition extracted from the present atmosphere models to those used by \ct{Saumon08}. We find that for our nominal $f_{\rm sed}=5$, $h=0.5$ partly cloudy sequence, the entropy at the bottom of the atmosphere (which gives the entropy of the matching interior model) is quite close to the cloudless case at $T_{\rm eff}=450\,$K. As the object cools the entropy decreases, and for $T_{\rm eff}=200\,$K the entropy is close to the entropy of cloudy atmospheres used in \ct{Saumon08}. This nicely corresponds to the transition of the partly cloudy models from optically thin to optically thick water clouds. Figures 4 and 5 of \ct{Saumon08} show that for given $T_{\rm eff}$ and $\log g$, the radii between the cloudless and cloudy evolution sequences vary by at most 1--2\% below 500$\,$K. Thus, the inconsistency between the surface boundary condition used in the evolution sequences and the model atmospheres presented here causes at most a 4\% error in the absolute fluxes.

\section{Results}

We present results for the grid of models discussed in Section \ref{modelgrid}. Where appropriate, we incorporate warmer models of T dwarfs from previous studies \cp{Saumon12, Morley12} for comparison. In Section \ref{cloudprops}, we present the model cloud properties. In Section \ref{pt-struc} we present the temperature structures of the models. In Section \ref{specs} we present the model spectra, including effects of disequilibrium chemistry. In Section \ref{colors} we present model photometry and compare to the growing collection of very cool objects with known distances \cp{Dupuy13, Beichman14}. Lastly, in Sections \ref{jwst} and \ref {gpi} we will make predictions for the characterizability of Y dwarfs with \jwst\ and the detectability of cool planets with new instruments like GPI, SPHERE, and the LBT. 

\subsection{Cloud properties} \label{cloudprops}
 \begin{figure}[t]
 \center    \includegraphics[width=3.75in]{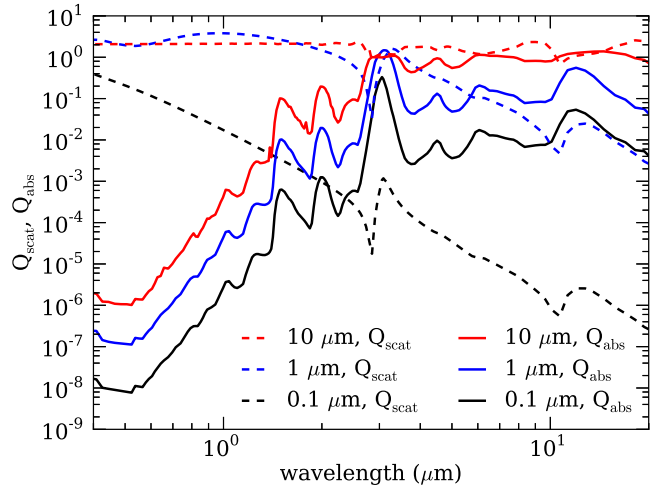}
  \caption{Absorption and scattering efficiencies. The results of the Mie scattering calculation (Q$_{\rm scat}$ and Q$_{\rm abs}$) for water clouds of three particle sizes are shown. These results are for single particle sizes, not a distribution of sizes. All three show similar general properties, with low Q$_{\rm abs}$ in the optical rising into the infrared and the strongest absorbing feature around 3 \micron. Larger particles are more efficient at both absorbing and scattering for most wavelengths. }
\label{miescattering}
\end{figure}

The models presented here include the effects of both sulfide/chloride clouds (first included in model atmospheres in \ct{Morley12}) and of water clouds. We will mainly focus on the properties of water clouds as the former set of clouds are more thoroughly examined in \ct{Morley12}. 

\subsubsection{Scattering and absorption efficiencies of water ice}
Figure \ref{miescattering} shows the optical properties calculated using Mie theory for water ice particles with sizes of 0.1, 1, and 10 \micron. The scattering efficiency, Q$_{\rm scat}$, is the ratio of the scattering cross section of the particle to the geometric cross section. The absorption efficiency, Q$_{\rm abs}$, is the ratio of the absorbing cross section to the geometric cross section. In general, the larger particle sizes both scatter and absorb more efficiently than smaller particles for most wavelength ranges. The locations of features are similar for different particle sizes with the strongest feature in Q$_{\rm abs}$ at 3 \micron. In general, Q$_{\rm abs}$ rises from optical to infrared wavelengths and remains fairly high through the infrared. The persistence of these features over a large range in particle sizes suggests that water ice features may be observable for a relatively optically thick cloud even if it contains a range of particle sizes. 

\begin{figure}[tb]
 \center    \includegraphics[width=3.5in]{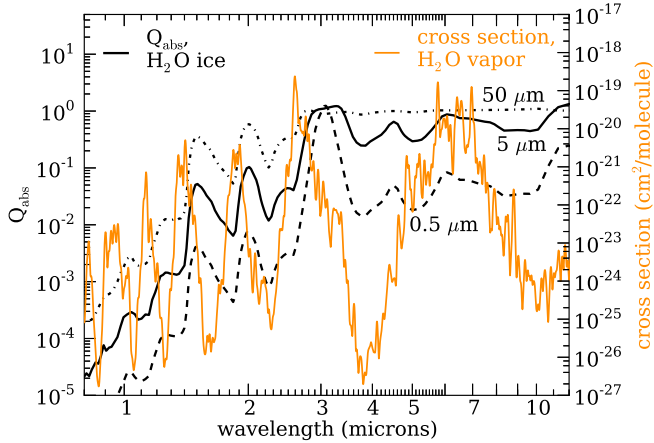}
  \caption{Absorption efficiency of water ice particles and absorption cross section of water vapor. The absorption efficiency Q$_{\rm abs}$ of water ice particles of three particle sizes (0.5, 5, and 50 \micron) is shown (left axis). These results are for single particle sizes, not a distribution of sizes. The absorption cross section of water vapor is shown on the right axis. The phase change of water substantially changes the wavelengths at which it strongly absorbs, filling in many of the regions where water vapor is transparent. }
\label{h2oabs}
\end{figure}

Figure \ref{h2oabs} shows both the absorption efficiency of water ice and the cross section of water vapor; the wavelength range at which water absorbs shifts as it condenses from vapor to solid phase. In particular, water ice absorbs strongly within the major water vapor opacity windows in the mid-infrared. 

\subsubsection{Particle sizes and optical depths of water clouds}

 \begin{figure}[t]
 \center    \includegraphics[width=3.75in]{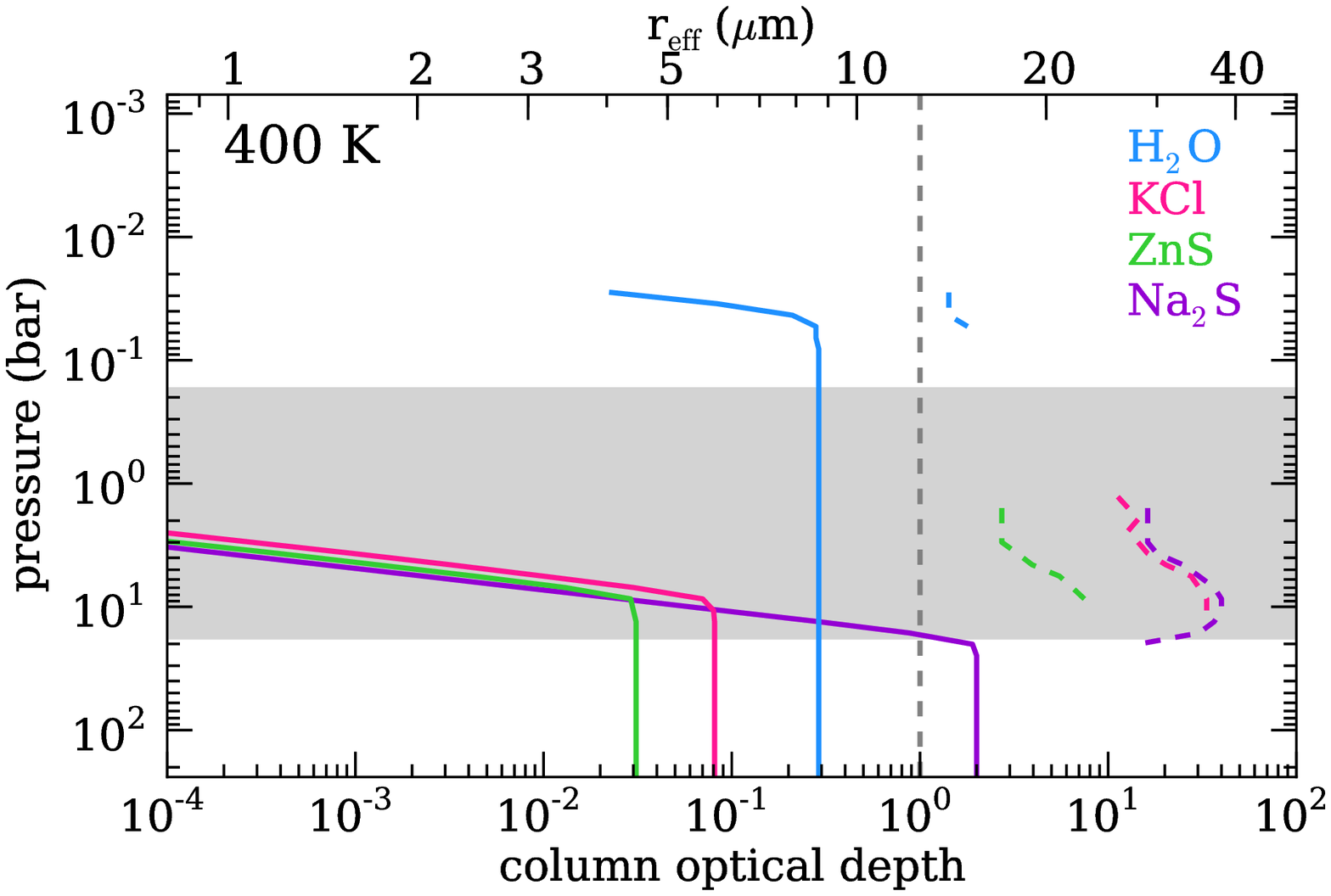}
\vspace{-0.3in}
 \center    \includegraphics[width=3.75in]{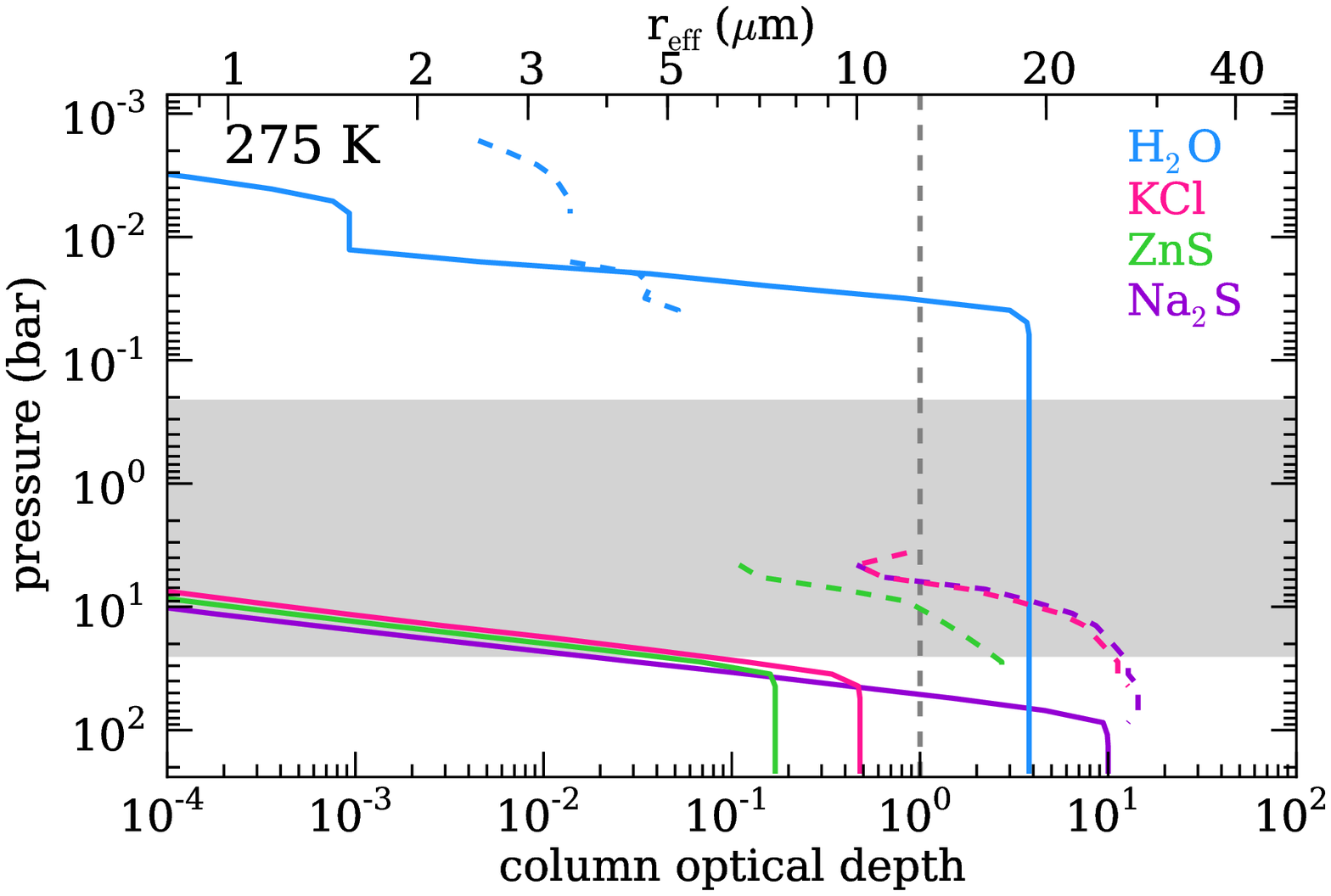}
\vspace{-0.3in}
 \center    \includegraphics[width=3.75in]{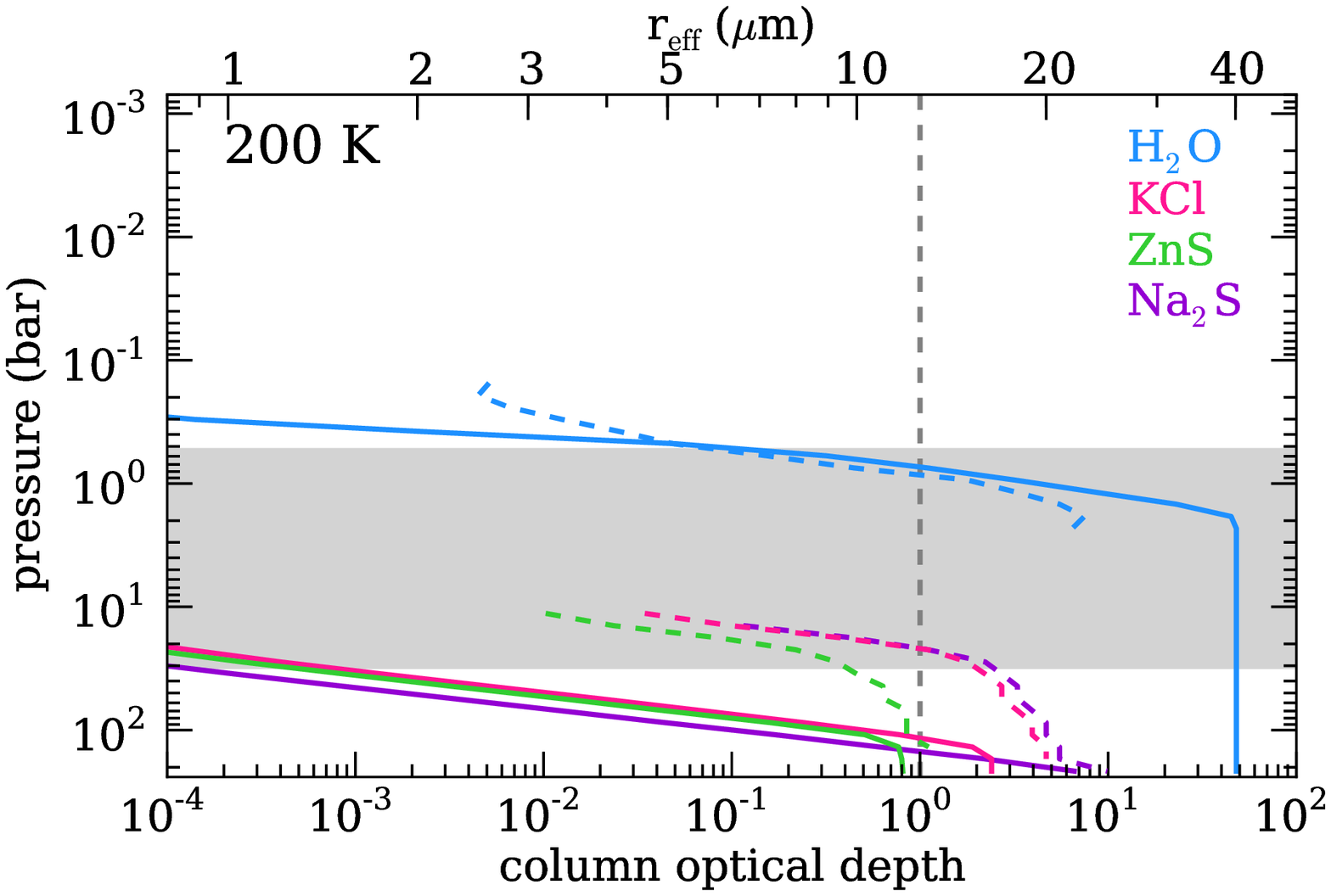}
  \caption{Cloud properties for sulfide/salt and water clouds at three temperatures. The geometric column optical depth is shown as solid lines. The effective (area-weighted) mode radius of the cloud particles at each pressure is shown as dashed lines. The 1--6 \micron\ photosphere is shown as the shaded gray region, and the $\tau=1$ line is shown to guide the eye. Thin water clouds form in all three models, but only become optically thick in the two coolest models. Mode particle sizes are small (3--5 \micron) for \teff=275 K and larger (5--20 \micron) for the 200 K model. The sulfide/salt clouds form and become optically thick in the photosphere of the 400 K model but are optically thick below the photospheres of the cooler two models as they form more deeply. }
\label{cloudcolopdreff}
\end{figure}

Water clouds form in brown dwarfs cooler than \teff=400 K, but initially in thin, tenuous layers. They first become relatively optically thick in the photospheres of brown dwarfs cooler than $\sim$350--375 K. Figure \ref{cloudcolopdreff} shows the cloud properties (mode particle size and geometric column optical depth) of a representative set of cloudy models. Geometric column optical depth is the equivalent optical depth of particles that scatter as geometric spheres; since water clouds are very much non-gray absorbers, this is a poor approximation at wavelengths where the particles scatter much more strongly than they absorb. 

\teff=400 K model atmospheres have sulfide and salt clouds in the photosphere and a thin water cloud in the upper atmosphere. The cloud properties for an example \teff=400 K, log g=4.5 model is shown in the upper panel of Figure \ref{cloudcolopdreff}. The water cloud does forms at a pressure level of 4$\times10^{-2}$ bar, in the upper atmosphere, in an optically thin layer. The sodium sulfide cloud becomes optically thick ($\tau=2$) at 20 bar, near the bottom of the photosphere. Mode particle sizes of this dominant \nas\ cloud are around 20--30 \micron. 

As a brown dwarf cools below \teff=400 K, the water cloud forms deeper in the atmosphere and more material is available to condense, making the water cloud much more optically thick. The middle panel shows a cooler model, \teff=275 K, in which the cloud has become geometrically optically thick near the top of the photosphere. Mode particle sizes in this high cloud layer are fairly small---around 1--5 \micron. The feature in the column optical depth around 10$^{-2}$ bar is caused by the fact that the pressure--temperature profile becomes warmer than the water condensation curve at that pressure. In this model, the sulfide and chloride clouds become optically thick much deeper in the atmosphere, around 100 bar, which is below the photosphere. 

For a brown dwarf that has cooled to \teff=200 K, the water cloud is very optically thick and forms within the photosphere. The bottom panel shows a \teff=200 K model; the base of the water cloud is at 2 bar and the column optical depth is $\sim$60. Mode particle sizes of water ice in the photosphere are 4--20 \micron. The cloud opacity is the dominant opacity source through cloudy columns of the model atmosphere. 
 
Overall, and in agreement with \ct{Marley99} and \ct{Burrows03b}, we see that water clouds begin to form with small particle sizes high in the atmospheres of objects around 400 K. They become marginally optically thick for objects cooler than 350--375 K. For very cold objects, $\sim$200--250 K, the water cloud is a dominant opacity source through cloudy columns of the atmosphere. 

 \begin{figure}[tb]
 \center    \includegraphics[width=3.5in]{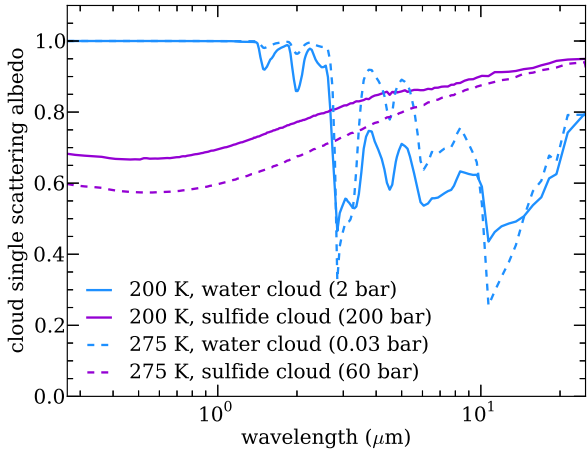}
  \caption{Single Scattering Albedo for water and \nas\ cloud. For models with \teff=200 K and 275 K, the single scattering albedos of both the water and \nas\ cloud are shown for a single atmospheric layer. The water cloud forms high in the atmosphere (2 bar and 0.03 bar for the layers shown from the 200 and 275 K models, respectively) and the \nas\ cloud forms deeper (200 and 60 bar, respectively). The sulfide cloud single scattering albedo is relatively uniform, rising from $\sim$0.6 in the optical to 0.9 at 10 \micron. The water cloud single scattering albedo has many more features, which vary with particle size (the mode particle size is $\sim$20 \micron\ for the 200 K model and $\sim$5 \micron\ for the 275 K model; the single scattering albedo is calculated for the distribution of particle sizes calculated using the cloud code). In the optical the single scattering albedo is 1.0, which means that the water clouds do not absorb efficiently at short wavelengths. }
\label{ssa}
\end{figure}

\subsubsection{Single scattering albedos}
However, even if a cloud is geometrically optically thick, depending on the optical properties of the absorbing and scattering particles, the cloud may not dramatically affect the spectrum. If the absorption efficiency is very low at a given wavelength, photons from layers below the cloud will have a very low probability of being absorbed by the cloud; thus the spectrum will appear essentially as it would in a cloud-free atmosphere at that wavelength.  

The single scattering albedo quantifies the importance of scattering and absorption by cloud particles. It is the ratio of the scattering coefficient to extinction coefficient (including both scattering and absorption) at a given wavelength. A single scattering albedo of 1.0 indicates that the cloud particles are entirely scatterers; a single scattering albedo of 0.0 indicates that the cloud particles are entirely absorptive. 

Figure \ref{ssa} shows the single scattering albedo for the clouds in the \teff=200 and 275 K models also shown in Figure \ref{cloudcolopdreff}. We show both the single scattering albedo for the sodium sulfide cloud, which is deep within the atmosphere at $\sim$ 100 bar, and the water ice cloud, which is in the photosphere around 0.1--1.0 bar. The single scattering albedo of the sulfide cloud is almost identical for each of these two model atmospheres; it rises from 0.6 in the optical to 0.95 at 12 \micron, indicating that the cloud becomes less efficient at absorbing as the wavelength increases. In contrast, the single scattering albedo of the water ice cloud has strong absorption features throughout the near and mid-infrared. The two models have slightly different features because the particle sizes are different, and the scattering and absorption properties depend fairly strongly on particle size (see Figure \ref{miescattering}). For the 200 K object, the water ice mode particle size at 2 bar is about 20 \micron. For the warmer 275 K model, the mode particle size at 0.03 bar is about a factor of four smaller, $\sim$5 \micron. 

However, the strongest features are evident for both models. The most obvious is the sharp decrease in single scattering albedo at 2.8 \micron, indicating that the cloud becomes more strongly absorbing at that wavelength. The feature at 10 \micron\ is very evident for the warmer model, and much more muted in the cooler model. The presence of features indicates that water ice is not a mostly gray absorber (like many of the more refractory clouds are) and, when present and optically thick, may cause spectral features, including ones that depend on particle size.

\subsection{Pressure--temperature structure} \label{pt-struc}

 \begin{figure}[tb]
 \center   \includegraphics[width=3.75in]{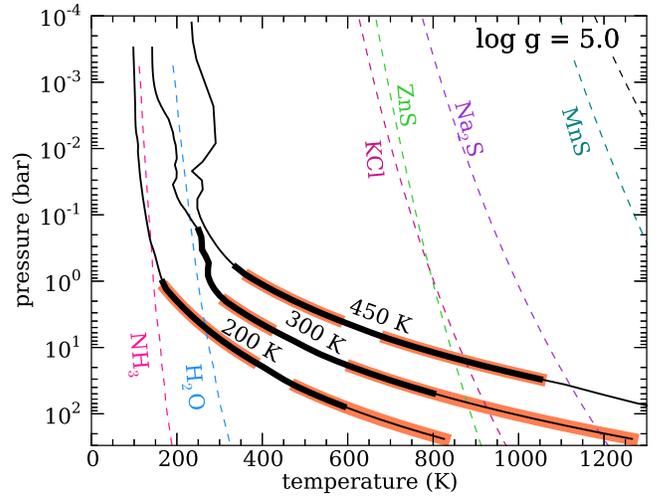}
\vspace{-0.5in}
 \center    \includegraphics[width=3.75in]{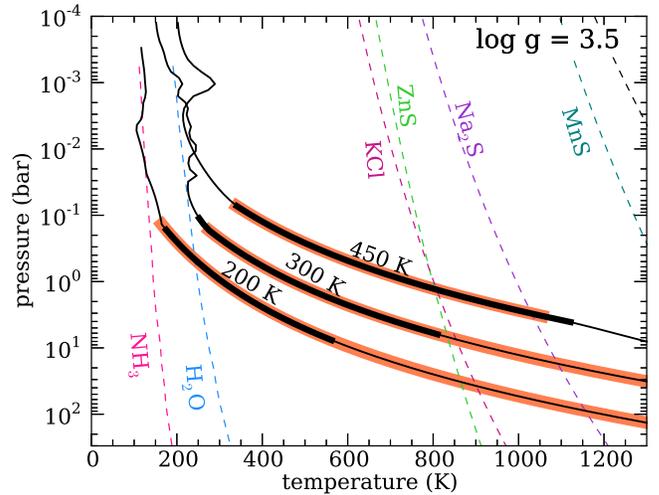}
  \caption{Pressure--temperature profiles for three representative temperature and two gravities are shown. The thicker black line indicates the location of the 1--6 \micron\ photosphere. The shaded salmon region shows where the atmosphere is convective. The dashed lines show condensation curves for each substance expected to condense in thermochemical equilibrium. The curve represents the pressure--temperature points at which the partial pressure of the gas is equal to the saturation vapor pressure; to the left of the curve, the partial pressure of each gas is higher than the saturation vapor pressure and the excess vapor will form a cloud. The kinks in the profile in the upper atmosphere are numerical and do not represent `real' features in the atmospheres of Y dwarfs. Fortunately, the kinks lie above the regions of the atmosphere from which flux emerges and so they do not pose a problem for this work. }
\label{pt-profile}
\end{figure}

Examples of model pressure--temperature profiles for the model grid are shown in Figure \ref{pt-profile}. In general, cloud opacity in a brown dwarf increases the temperature of the atmosphere at all points in the atmosphere. This is because in a cloudy atmosphere, the overall opacity is slightly higher, so the temperature structure of a converged model with the same outgoing flux will be slightly warmer. 

For a cloud-free model atmosphere, once the water has condensed out of the atmosphere, there are very few opacity sources left: mainly CH$_4$, NH$_3$, and collision-induced absorption from molecular hydrogen. This means that the brown dwarf is quite transparent in those layers and flux is able to emerge from deeper layers. In contrast, if we assume, as we do in our cloudy models, that water in excess of the saturation vapor pressure condenses to form a water ice cloud, that cloud provides a large opacity source, preventing the brown dwarf from efficiently emitting from layers underneath the cloud, and significantly warming the atmosphere. 

For model atmospheres between 400 and 500 K, even though the cloud-free P--T profile may cross the water condensation curve, the converged cloudy models (including the effect of sulfide clouds) are somewhat warmer and do not cross the water condensation curve, so the model water cloud does not form. 

For the warmest effective temperature brown dwarfs in which we find that water clouds form and can exist in radiative equilibrium, ($\lesssim$400 K),  the water clouds are at low pressures within the radiative upper atmosphere. They remain above the photosphere and optically thin. This means that the clouds affect the spectra very little, but the converged cloudy model's P--T profile is warmer by $\sim$20-50 K in the upper atmosphere than a corresponding cloud-free model. 

Figure \ref{pt-profile} shows the models of pressure--temperature profiles of brown dwarfs (log g=5.0, upper panel) and planet-mass objects (log g=3.5, lower panel), with effective temperatures of 200, 300, and 450 K. The photospheres, shown as thick black lines, show that the observable region of the atmospheres tends to be below the radiative upper atmospheres which are prone to numerical challenges in the models. The convective regions are also shown as colored shaded regions; higher gravity objects at these temperatures have multiple convection zones. This well-known result is because their P--T profiles cross regions of parameter space where the opacity of a solar composition equilibrium gas is low, and radiative energy transport becomes efficient. In lower pressure regions in the atmosphere, the opacity increases and radiative transport is once again inefficient and the model becomes unstable to convection \cp[e.g.][]{Marley96, Burrows97}.

Note that the coolest models (\teff=200 K) also cross the NH$_3$ condensation curve, indicating that for very cold Y dwarfs we will also need to consider the effects of the ammonia cloud. Like the H$_2$O cloud, it will first form as a thin cloud high in the atmosphere, and become more optically thick as the object cools further. The opacity of this cloud is not currently included in the models. 

\subsection{Spectra} \label{specs}

The spectra of Y dwarfs are dramatically different from blackbodies at the same effective temperatures, with strong molecular absorption features where the thermal emission would peak and opacity windows at shorter wavelengths where a blackbody would be faint. A 450 K object is 6 orders of magnitude brighter in J band than its blackbody counterpart; a 300 K object is 10 orders of magnitude brighter; a 200 K object is 15 orders of magnitude brighter. 

\subsubsection{Molecular absorption bands}

 \begin{figure*}[tb]
 \center    \includegraphics[width=6in]{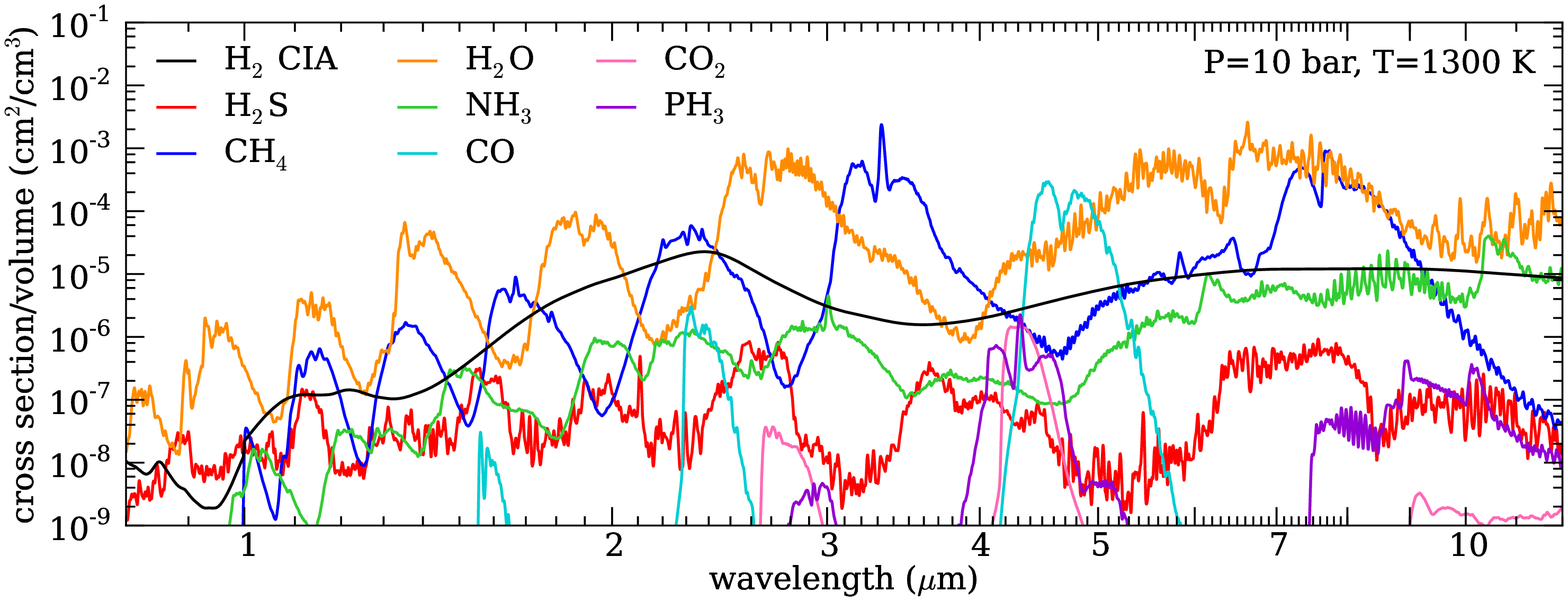}
\vspace{-0.35in}
 \center    \includegraphics[width=6in]{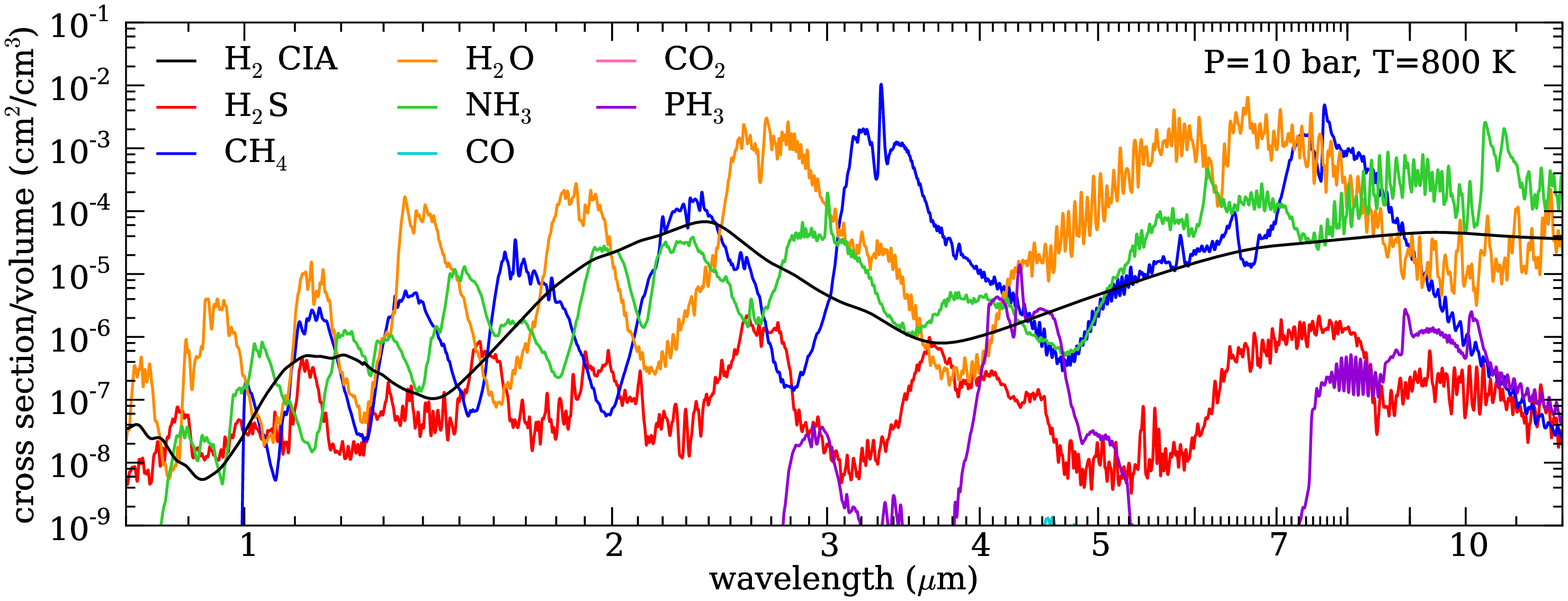}
\vspace{-0.35in}
 \center    \includegraphics[width=6in]{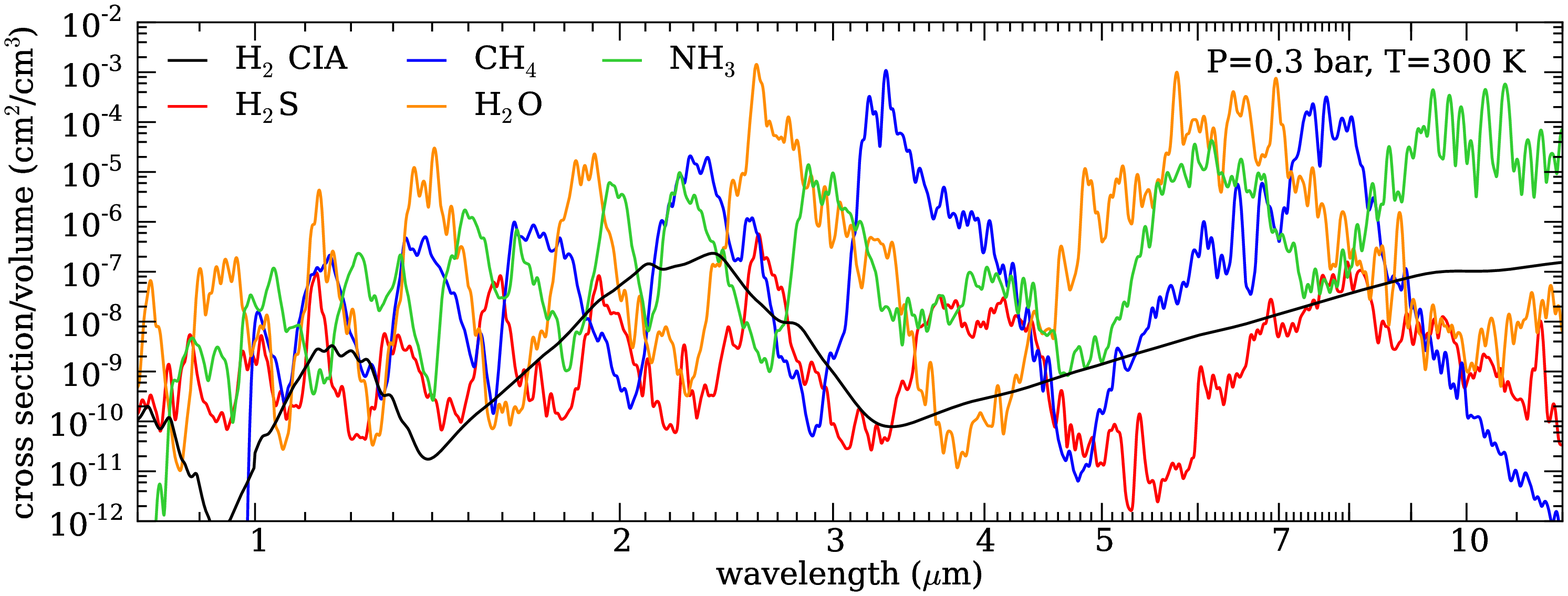}
\vspace{-0.35in}
 \center    \includegraphics[width=6in]{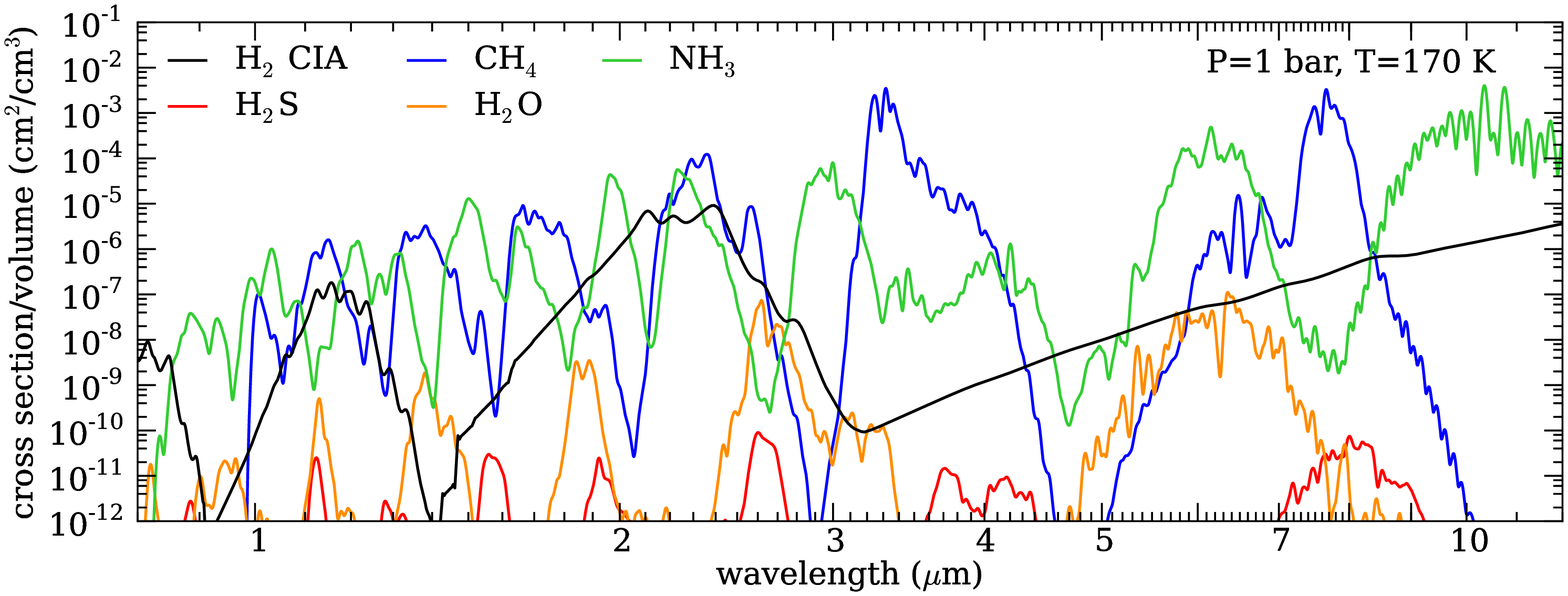}
  \caption{Opacities of the major constituents of Y and T dwarfs. We choose four representative P--T points in the photospheres of models at three different temperatures (all with log g=5.0): \teff=900 K (P=10 bar, T=1300 K), \teff=450 K (P=10 bar, T=800 K and P=0.3 bar, T=300 K), and \teff=200 K (P=1 bar, T=170 K). We multiply the molecular opacities (cm$^2$/molecule) by the number density of that molecule in a solar metallicity atmosphere in thermochemical equilibrium to get a opacity per volume of atmosphere. In this temperature range, the abundances of CO and CO$_2$ drop by orders of magnitude. Water vapor remains an important opacity source in the top three panels, but drops significantly in the bottom panel because of water condensation. NH$_3$ and CH$_4$ gradually become more important as objects cool. PH$_3$ may also be an important absorber for the Y dwarfs in the mid-infrared.   }
\label{molopac}
\end{figure*}

The spectra of Y dwarfs are dominated by the opacity of H$_2$O, CH$_4$, NH$_3$, and H$_2$ collision-induced absorption (CIA). Figure \ref{molopac} shows the molecular opacity and collision induced absorption at representative locations in the photospheres of objects with effective temperatures of 900 (T6.5), 450 (Y0), and 200 K (Y2+). This progression from mid-T to late-Y is marked by the increase in ammonia absorption relative to water and methane. 

 \begin{figure*}[t]
 \center    \includegraphics[width=7in]{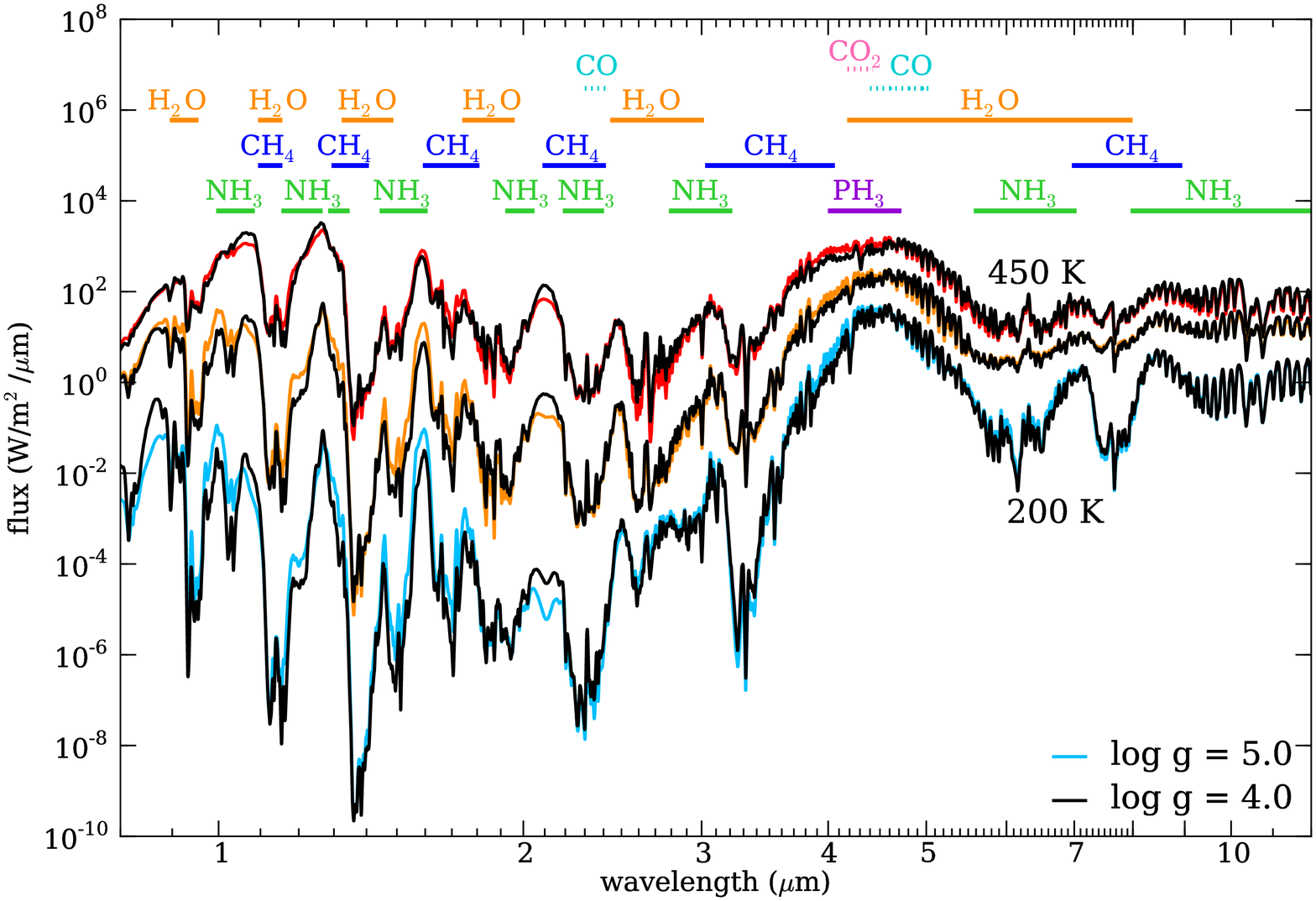}
  \caption{Model spectra of three effective temperature (450, 300, 200 K) at two gravities (log g=4.0, 5.0) and cloud parameters \fsed=5, $h$=0.5. Locations where each of the major molecules in the atmosphere peak in absorption are marked by the bands along the top. The near- and mid-infrared are carved by overlapping bands of water, methane, and ammonia absorption. The mid-infrared is also affected by PH$_3$.   }
\label{spec-log-bands}
\end{figure*}

Features from CO and CO$_2$ have been found in T dwarfs \cp{Yamamura10, Tsuji11}. The strongest band of CO is the dominant opacity source at 4.5--5 \micron, even assuming equilibrium chemistry, for a mid-T dwarf. CO$_2$ is also an important opacity source in the mid-infrared for warmer objects, especially if its mixing ratio is increased by vertical mixing. As an object cools, CO and CO$_2$ are strongly disfavored in equilibrium, but vertical mixing could increase their abundance in the atmosphere by several orders of magnitude, so the strongest absorption bands may still prove to be important for late T and early Y dwarfs. For very cold objects, the effects of disequilibrium carbon chemistry should become less important. 

Species such as PH$_3$ (phosphine) and H$_2$S, which have been observed in Jupiter's atmosphere \cp{Prinn75}, will also be present in Y dwarfs. Phosphine is likely observable in the mid-infrared; the strongest PH$_3$ feature is at 4.3 \micron\ and is the dominant opacity source at that wavelength in the photosphere of a \teff=450 K Y dwarf. While equilibrium models find little phosphine in the photospheres of \teff=200 K objects, \ct{Visscher06} predict that phosphine will be in disequilibrium in giant planet and T dwarf atmospheres and could be orders of magnitude more abundant than equilibrium models predict. The effect of phosphine on Y dwarf spectra may therefore be underestimated in these models and even more pronounced in real Y dwarfs. H$_2$S affects the spectra mostly in \emph{H} band, where it acts largely as a continuum absorber, depressing the \emph{H} band peak, and to a smaller degree the \emph{Y} band peak.

\begin{figure}[h]
 \center    \includegraphics[width=3.5in]{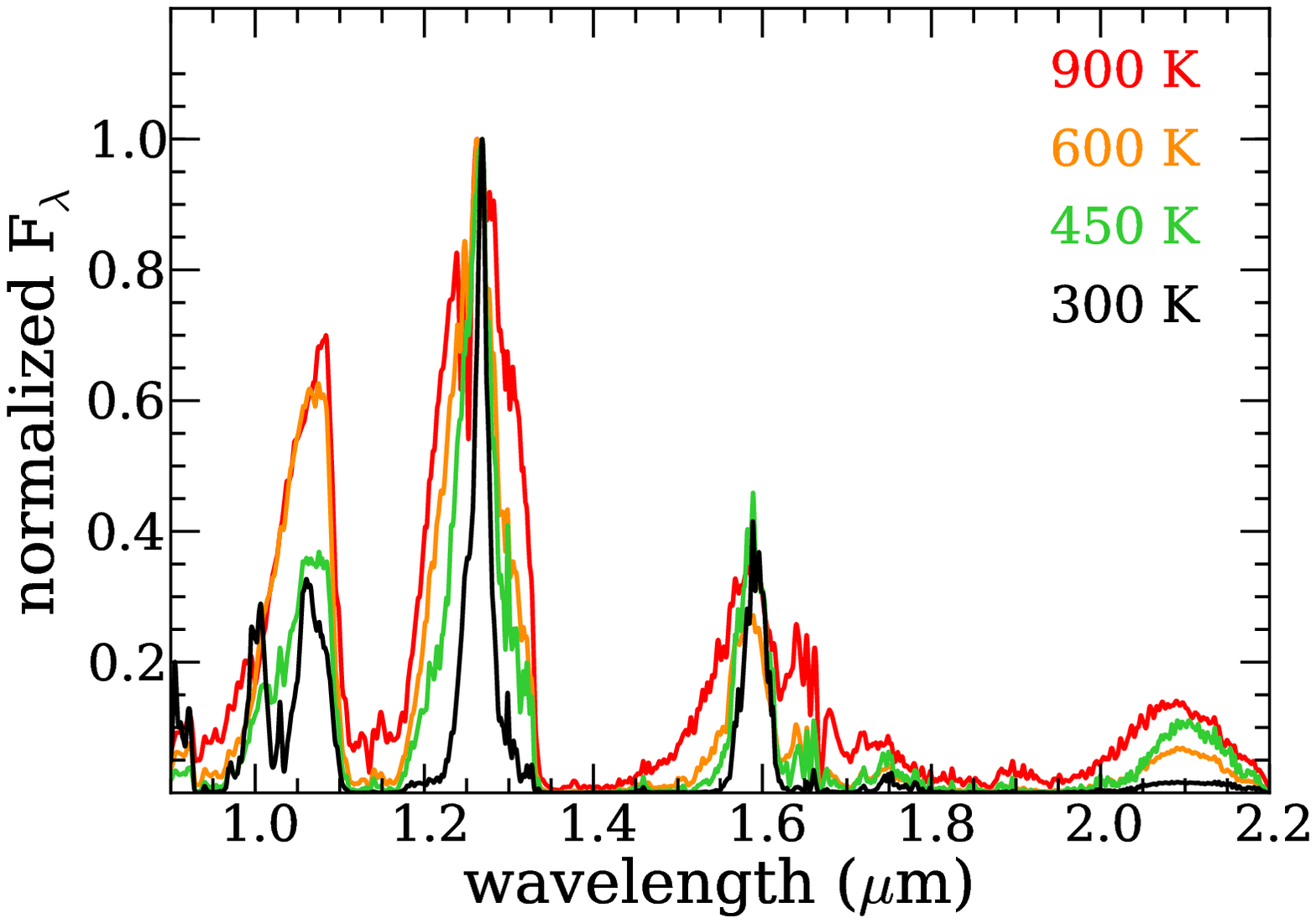}
  \caption{Model spectra at four effective temperature spanning mid-T to Y dwarfs (900, 600, 450, 300 K),  log g=4.5, and cloud parameters \fsed=5, $h$=0 (900, 600 K) and $h$=0.5 (450, 300 K). Spectra are rescaled such that the flux at the peak of \emph{J} band is the same for all models. Note the change in the shape of the near-IR spectral windows. \emph{J} and \emph{H} bands narrow as ammonia and methane increase in abundance. Ammonia absorption in \emph{Y} band causes the band shape to bifurcate for the coolest model. }
\label{irspec-zoom}
\end{figure}

 \begin{figure*}[th]
 \center    \includegraphics[width=7in]{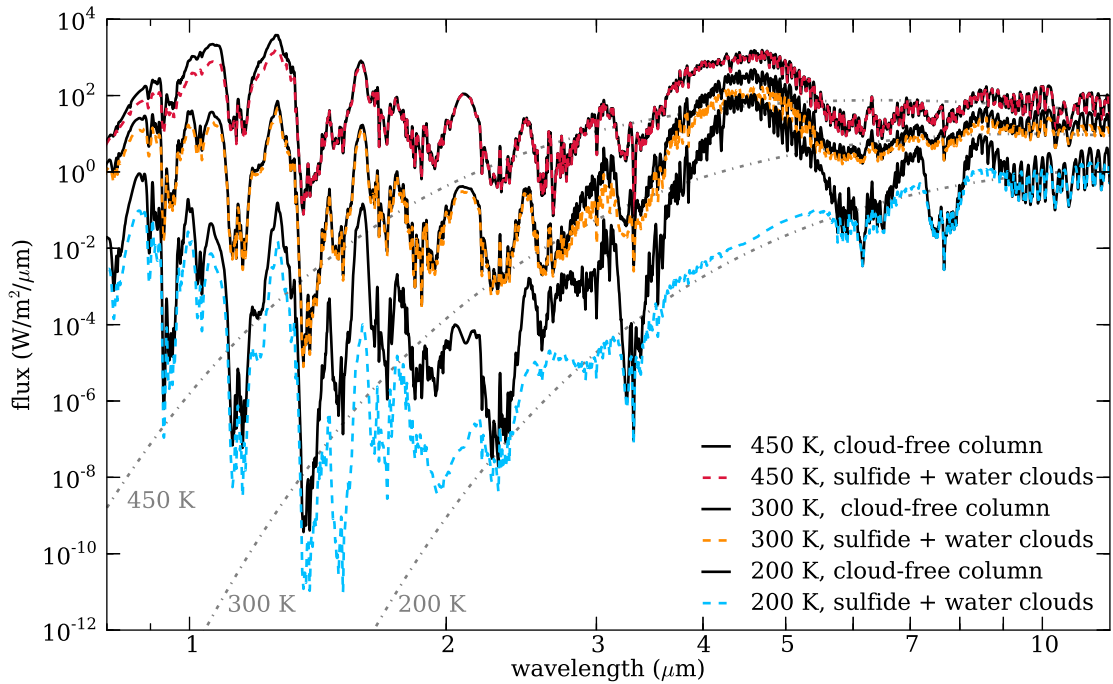}
  \caption{Clear and cloudy spectra of models of three effective temperature (450, 300, 200 K) with log g=5.0 and cloud parameters \fsed=5, $h$=0.5. Blackbodies of equivalent effective temperatures are shown as dashed gray lines. Each of the models shown for a given temperature has the same P--T profile; the cloud-free spectrum is the flux calculated through the clear column and the cloudy spectrum is the flux calculated through the cloudy column. Summed together, they have the correct effective temperature. More flux is able to emerge from the clear column because the opacity is lower. For the 450 K model, the greatest flux difference between the cloud-free and cloudy models is in \emph{Y} and \emph{J} bands. In the 300 K model, the greatest flux difference is at the flux peak at 4.5 \micron\ where the water clouds absorb strongly. For the 200 K model, the water cloud is very optically thick and within the photosphere, so at all the wavelengths where the water cloud absorbs, the flux emerging from the cloudy column is significantly limited.  }
\label{cloudyspecs}
\end{figure*}

 \begin{figure*}[th]
 \center    \includegraphics[width=6in]{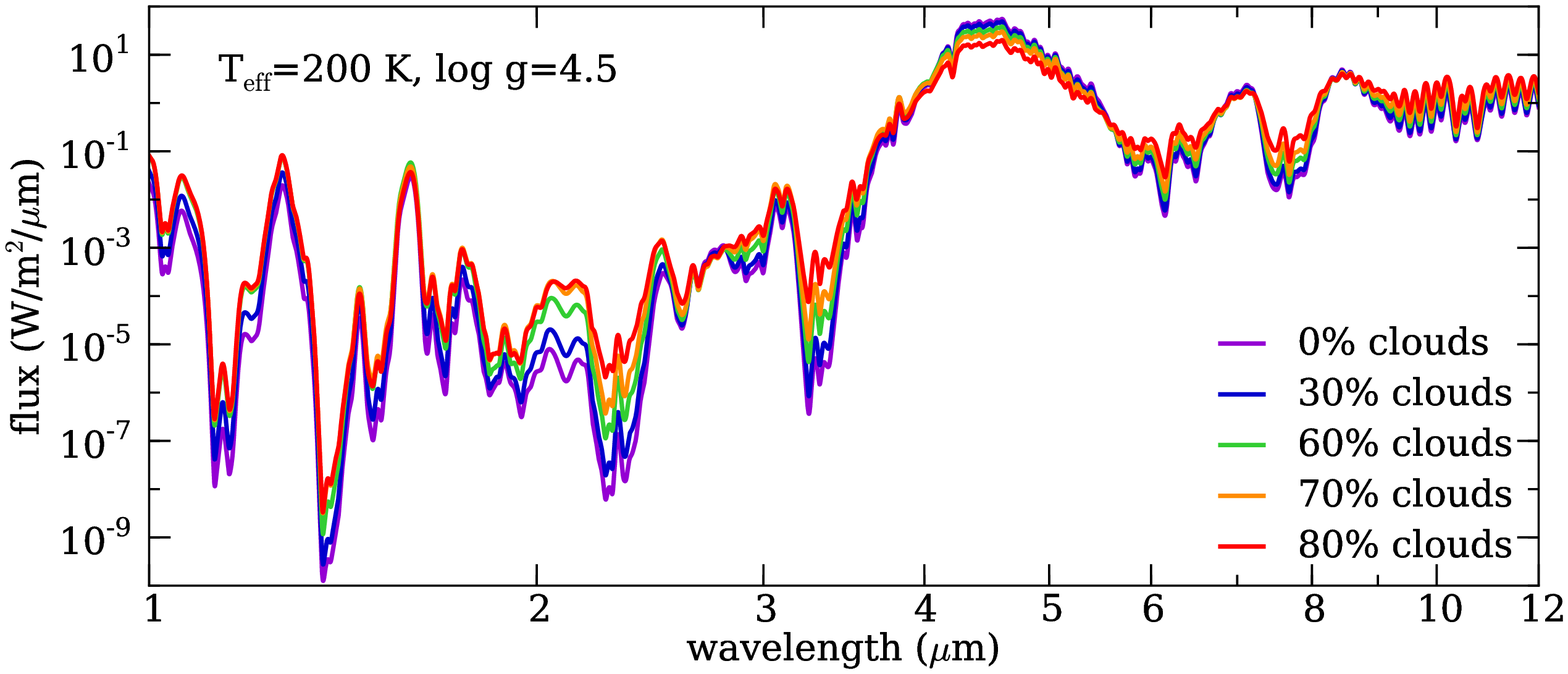}
 \vspace{-0.55cm}
 \center    \includegraphics[width=6in]{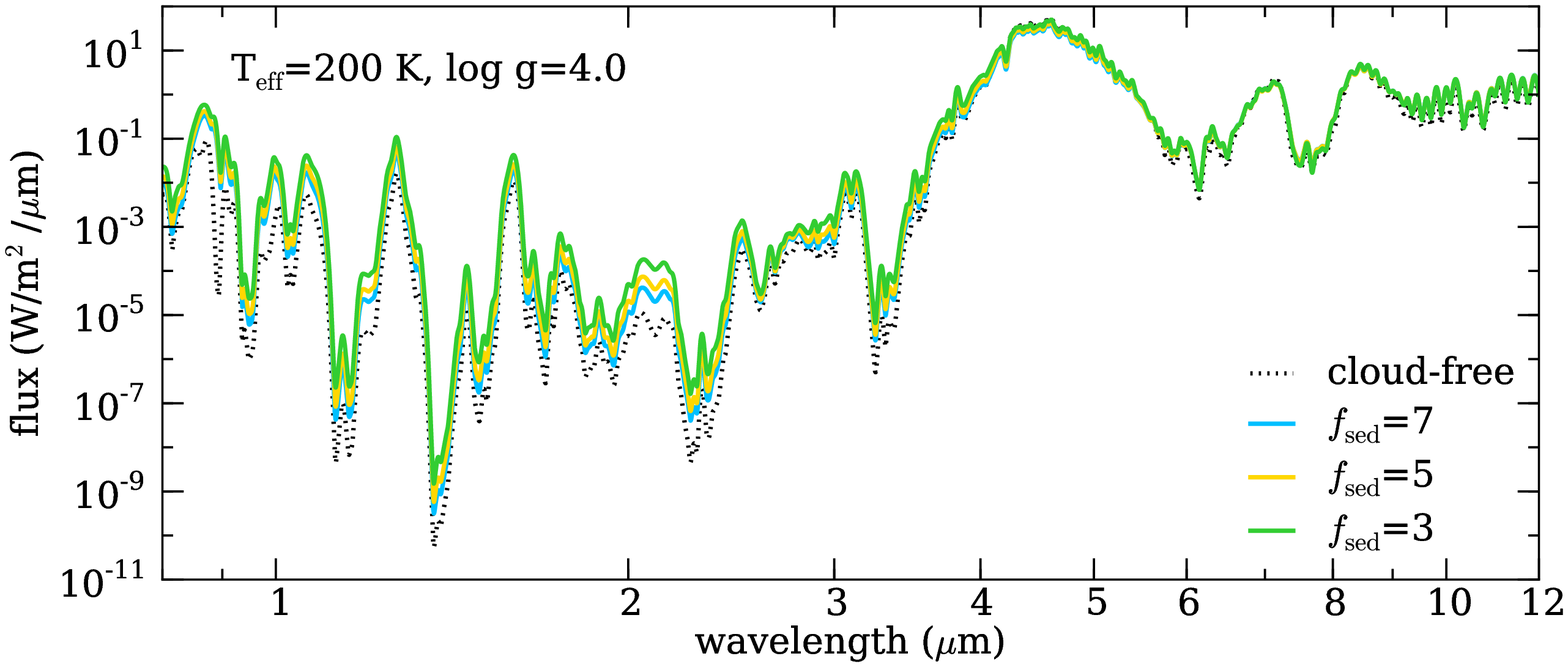} 
 \caption{Spectra of models in which we vary the two free parameters of the patchy cloud model, $h$ and \fsed. All the models shown have \teff=200 K. In the upper panel, $h$ is varied from 1.0 (cloud-free) to 0.2 (80\% of the surface covered in clouds) and \fsed=5. In the lower panel, \fsed\ is varied from 3 to 7 and $h$=0.5. The flux is redistributed when an atmosphere is cloudy; all models have the same total amount of energy emerging. Most prominently, clouds decrease the flux in the major flux peak at 4--5 \micron\ and redistribute that energy from the flux peak into other parts of the spectrum. For example, the cloudiest model is significantly brighter at the \emph{K} band peak than the cloud-free model. }
\label{diffcloudcover}
\end{figure*}

\subsubsection{Model spectra}

Model spectra of objects at two different representative gravities (log g=5.0, 4.0) from \teff=450 to 200 K are shown in Figure \ref{spec-log-bands}. The models shown assume \fsed=5 and $h=0.5$, as described in Section \ref{partlycloudymodels}, and include both the salt/sulfide clouds (\nas, KCl, ZnS, MnS, Cr) and water ice clouds. As a brown dwarf cools over the Y dwarf sequence, the near-infrared flux dramatically declines. By the time the object has cooled to 200 K, almost all flux emerges in the mid-infrared, between strong molecular absorption features. 

Figure \ref{spec-log-bands} also shows the locations of the dominant absorption bands. For objects at these temperatures, absorption is dominated by water, methane, and ammonia, with contributions from PH$_3$ in the mid-infrared, including the feature at 4.3 \micron\ for the 450 K model spectrum. Features from CO and CO$_2$ are not present here due to the low mixing ratios of these molecules at these temperatures in chemical equilibrium. Minor differences exist between models at the two shown gravities. We explore the gravity-dependence of features more in the discussion of Figure \ref{gravsigs}. 

Figure \ref{irspec-zoom} shows how the changes in molecular abundances and absorption changes the shape of the near-infrared spectra over the T to Y sequence. The wide spectral windows in the water absorption, typical of warmer brown dwarfs, narrow as ammonia and methane increase in abundance in the near-infrared photosphere. As ammonia increases in abundance, the weaker ammonia feature between 1 and 1.1 \micron\ begins to carve away the center of \emph{Y} band. For the 300 K model shown in Figure \ref{irspec-zoom}, the \emph{Y} band is bifurcated into two peaks by this absorption band. The appearance of this split \emph{Y} band will depend on the nitrogen chemistry; if ammonia is less abundant due to disequilibrium chemistry, this change may occur at a lower temperature. The decline of the alkali absorption with temperature (see Figure \ref{alkalis}) will also affect the underlying continuum absorption in \emph{Y} band and therefore the appearance of this split.

\subsubsection{The effect of sulfide and water clouds}
Figure \ref{spec-log-bands} shows the summed flux through the cloudy and clear columns of the model atmosphere. However, the flux through each of those columns is not equal; since the opacity of the cloud increases the total opacity through the column, less flux always emerges through the cloudy column than through the clear column. 

Figure \ref{cloudyspecs} shows models at the same effective temperatures, now showing the flux from the cloudy and clear columns. At effective temperatures above 400 K, the water cloud has not yet formed or is extremely thin, so the cloudy and cloud-free columns look quite similar; they differ only substantially in the $Y$ and $J$ bands, which is the region that the deep sulfide clouds affect. At effective temperatures between 300--375 K, the water cloud gradually becomes more optically thick. It first forms quite high in the atmosphere, and influences the mid-infrared from 2.8--5 \micron\ where water ice particles absorb most efficiently (see Figure \ref{ssa}). 

For cold Y dwarfs---between 200-300 K---the water cloud thickens as the object cools. At effective temperatures of 200 K, it has become quite optically thick: most regions of the near- and mid-infrared have significantly less flux emerging from the cloudy column. In fact, about 10$^4$ times less flux emerges at 4.5 \micron\ from the cloudy column than the cloud-free column. This picture, where flux is emitted almost entirely through clearer columns of the atmosphere, is similar to Jupiter and Saturn's deep water clouds, which appear to have holes in the clouds (the so-called `5-micron hot spots' in Jupiter) through which most of the mid-infrared flux emerges. 

Interestingly, water ice particles of this size (1--20 \micron) are very inefficient at absorbing photons with wavelengths shorter than 1.4 \micron, so the $J$ and $Y$ bands are not strongly affected, even in a column with a geometrically optically thick cloud. 

It is also instructive to look at models that have the same total amount of flux emitted through model atmospheres with different cloud-covering fraction $h$ and different sedimentation efficiency \fsed, to understand the sensitivity of our results to our choice of those parameters. The models presented in Figure \ref{diffcloudcover} are separate from our main grid, which was run with $h=0.5$ and \fsed=5 for all models; these models instead are run with $h$=1.0, 0.7, 0.4, 0.3, and 0.2 and with \fsed=3, 5, and 7 respectively. All models have the same amount of total flux emitted as a 200 K blackbody. 

In both panels, the models look very similar to each other at wavelengths shorter than about 2 \micron, where the water clouds do not strongly absorb. As we increase the cloud fraction (decrease $h$), more flux emerges between 2 and 3.6$\,\mu$m and beyond 5.5$\,\mu$m. This additional flux comes at the expense of the peak flux at $\sim4.5\,\mu$m. In essence, increases in cloudiness redistribute this peak flux to other wavelengths. When we vary \fsed, as expected, lower values of \fsed\ have somewhat more cloud opacity; this effect is largest in \emph{K} band.

\subsubsection{Disequilibrium chemistry}

Disequilibrium carbon chemistry was predicted by \ct{Fegley96} and confirmed in spectra by \ct{Noll97} and \ct{Saumon00}. It is known to be important for brown dwarfs of many temperatures \cp{Saumon06, Hubeny07} and may be even more significant for young planets \cp{Konopacky13}. The atmospheres of cool brown dwarfs should be methane-dominated. However, the chemical reaction that leads to methane formation is strongly temperature sensitive and becomes very slow at cold temperatures. If the timescale of mixing in the atmosphere is faster than the timescale for this reaction to occur, carbon will remain in CO instead of being converted to the CH$_4$ favored by equilibrium chemistry; for very cold objects this will become less important as the region where methane is thermochemically favored will extend very deeply into the atmosphere. This process has been explored in detail in a number of papers \cp{Saumon06, Hubeny07, Visscher11, Moses11}.  

Disequilibrium chemistry also affects other abundant molecules, such as the conversion of N$_2$ to NH$_3$ and CO$_2$ to CH$_4$. Other elements such as phosphorous are also out of chemical equilibrium in cold planets like Jupiter; in equilibrium, phosphorous would be in the form P$_4$O$_6$ but it is instead observed as PH$_3$ because the pathways for forming P$_4$O$_6$ are kinetically inhibited \cp{Visscher06}. 

Here, we use the approach developed in \ct{Smith98} and techniques presented in \ct{Saumon06} to approximate the effect of CO/CH$_4$ and N$_2$/NH$_3$ disequilibrium in the atmospheres of Y dwarfs. Using this approach, we calculate a quench point in the atmosphere where the mixing timescale is equal to the chemical reaction timescale, which is governed by the slowest step of the fastest pathway of the reaction.  Above the quench point, we assume that the mixing ratio of the molecule is constant. This has been shown using full kinetics models to be a good approximation in substellar atmospheres \cp{Visscher11}. Note that in the disequilibrium chemistry calculation we calculate \kzz\ in the convective zone using mixing length theory and vary \kzz\ in the radiative zone as a free parameter, between 10$^2$--10$^6$ cm$^2$/s. In contrast, in the cloud code, we calculate \kzz\ using mixing length theory with a minimum \kzz\ of 10$^5$ cm$^2$/s; the clouds and chemistry are thus not strictly self-consistent in the radiative region. 

\begin{figure*}[htb]
 \center    \includegraphics[width=7in]{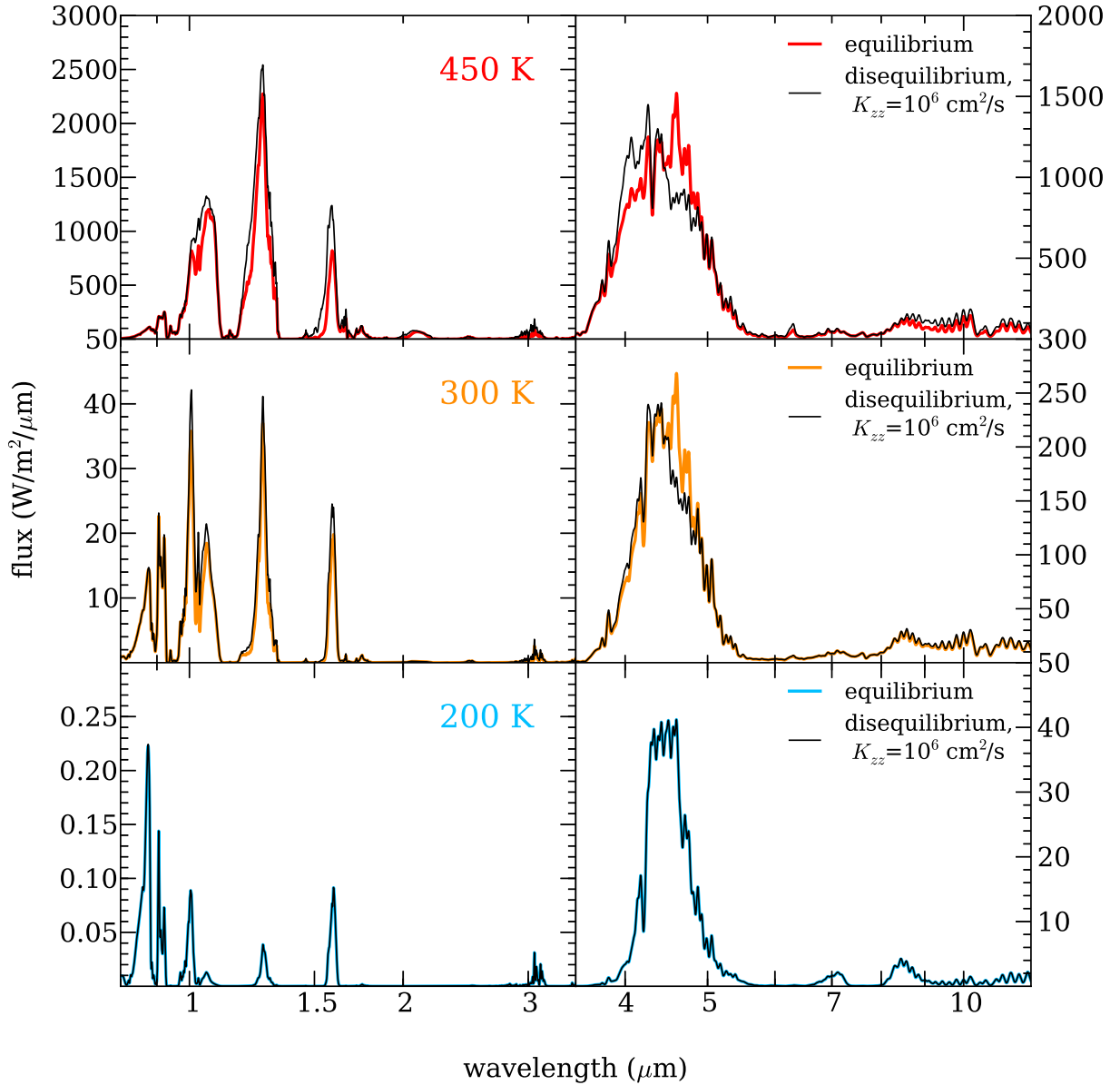}
  \caption{Spectra of models including disequilibrium chemistry at \teff=450, 300, and 200 K and log g=5.0. All disequilibrium models use eddy diffusion coefficient \kzz=10$^6$ cm$^2$/s and include CO/CH$_4$ and N$_2$/NH$_3$ disequilibrium. Near- and mid-infrared spectra are shown on axes with different linear scales to facilitate viewing small changes in spectra. At 450 K, in disequilibrium slightly more flux emerges from \emph{Y}, \emph{J}, and \emph{H} bands, the shape of the 4.5 \micron\ peak changes, and slightly more flux emerges from 8--12 \micron. At 300 K, the equilibrium and disequilibrium models do not differ as strongly, though the shape of the 4.5 \micron\ peak changes. At 200 K, the equilibrium and disequilibrium models are indistinguishable. }
\label{diseq}
\end{figure*}

The results of these calculations are shown in Figure \ref{diseq} for three test cases. For the 450 K model, the disequilibrium model is generally slightly brighter across the near-infrared than the equilibrium model. This is because in equilibrium, ammonia is strongly thermochemically favored; out of equilibrium, there is slightly more N$_2$, which is not a strong absorber, and slightly less NH$_3$, which absorbs strongly across the near-infrared (see Figure \ref{molopac}). In the mid-infrared, the disequilibrium model is slightly brighter around 4 \micron\ and slightly fainter around 4.6 \micron. This is due to the increase in CO and decrease in CH$_4$; this shift increases absorption from the most prominent CO band at 4.5 to 4.9 \micron\ and decreases absorption from both CH$_4$ and NH$_3$ between 3 and 4.4 \micron. The increase in flux beyond 8 \micron\ is because of the decrease in NH$_3$ which is the strongest absorber at those wavelengths. 

For the 300 K and 200 K objects shown in Figure \ref{diseq}, disequilibrium chemistry of these particular gases becomes less important as the objects cool. The atmospheres of these colder objects favor CH$_4$ and NH$_3$ strongly in equilibrium over a progressively wider proportion of their atmospheres. This means that even if deeper layers were mixed upwards to the photosphere, those layers would still be dominated by CH$_4$ and NH$_3$. 

\begin{figure}[htb]
 \center    \includegraphics[width=3.75in]{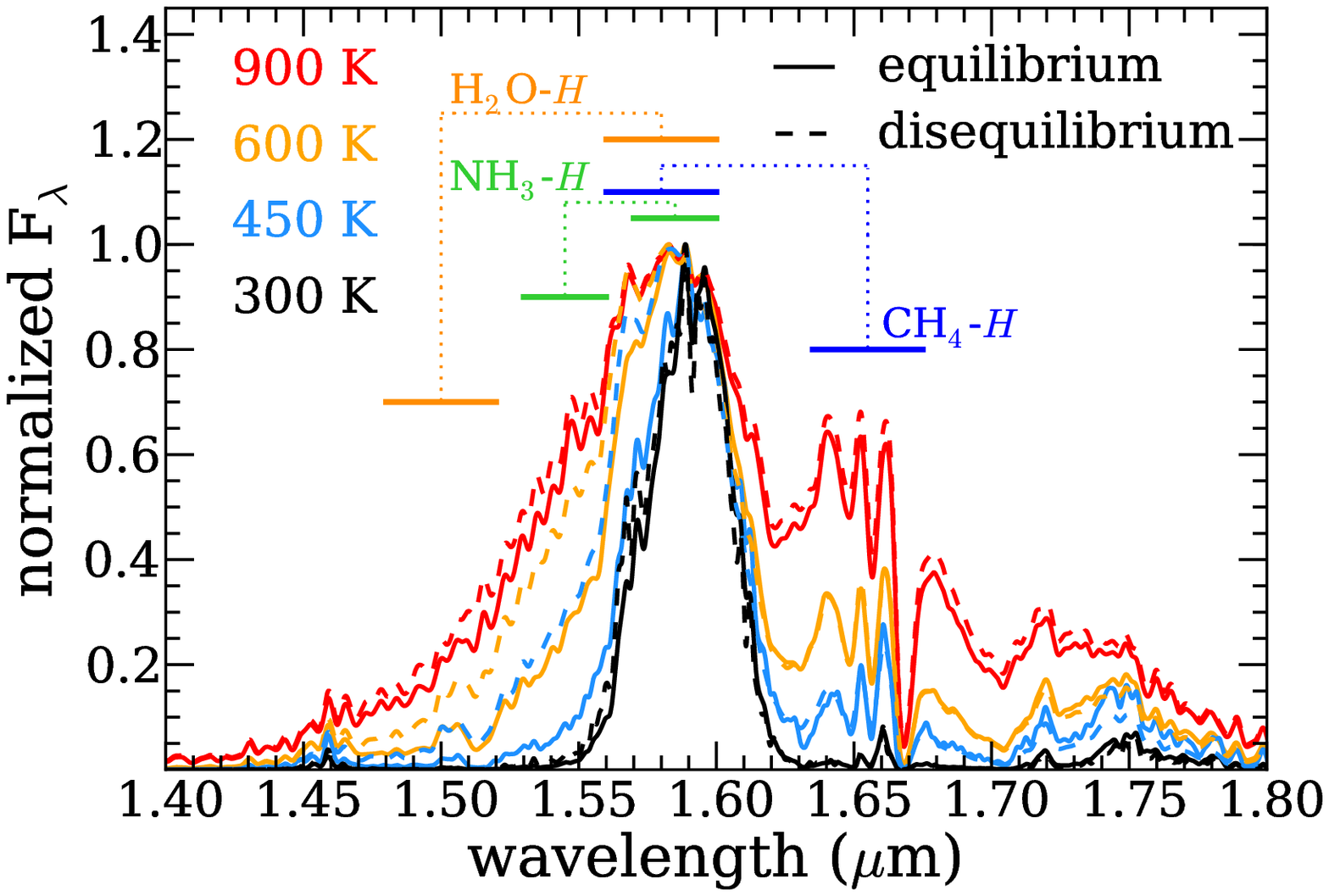}
  \caption{Shape of the \emph{H} band over the late T to Y sequence. As ammonia and methane absorption on the blue and red sides of the \emph{H} band, the peak narrows. The disequilibrium models (\kzz=10$^4$ cm$^2$/s) narrow more slowly on the blue side where ammonia absorbs because disequilibrium chemistry decreases the amount of NH$_3$ and increases the amount of N$_2$. The locations of spectral indices used to classify T dwarfs are shown \cp{Burgasser06a, Delorme08}.}
\label{hbandammonia}
\end{figure}

The shape of the $H$ band has been used as an indicator of the increased abundance in ammonia through the T sequence to the Y dwarfs as they cool (see Figure 5 from \ct{Cushing11}). We show a modeler's version of the same sequence in Figure \ref{hbandammonia} with the three sets of spectral indices used to classify T dwarfs \cp{Burgasser06a, Delorme08}. The blue side of $H$ band narrows as the object cools, due largely to increased ammonia absorption from 1.5 to 1.6 \micron. In equilibrium, the shape changes from 900--450 K as the ammonia absorption increases. Disequilibrium chemistry changes the progression somewhat because the ammonia appears more gradually from 900--300 K. The shape of the disequilibrium $H$ band at 450 K is very similar to the shape of the equilibrium $H$ band at 600 K. 

\subsubsection{Decline in the alkali absorption}

 \begin{figure}[htb]
 \center    \includegraphics[width=3.75in]{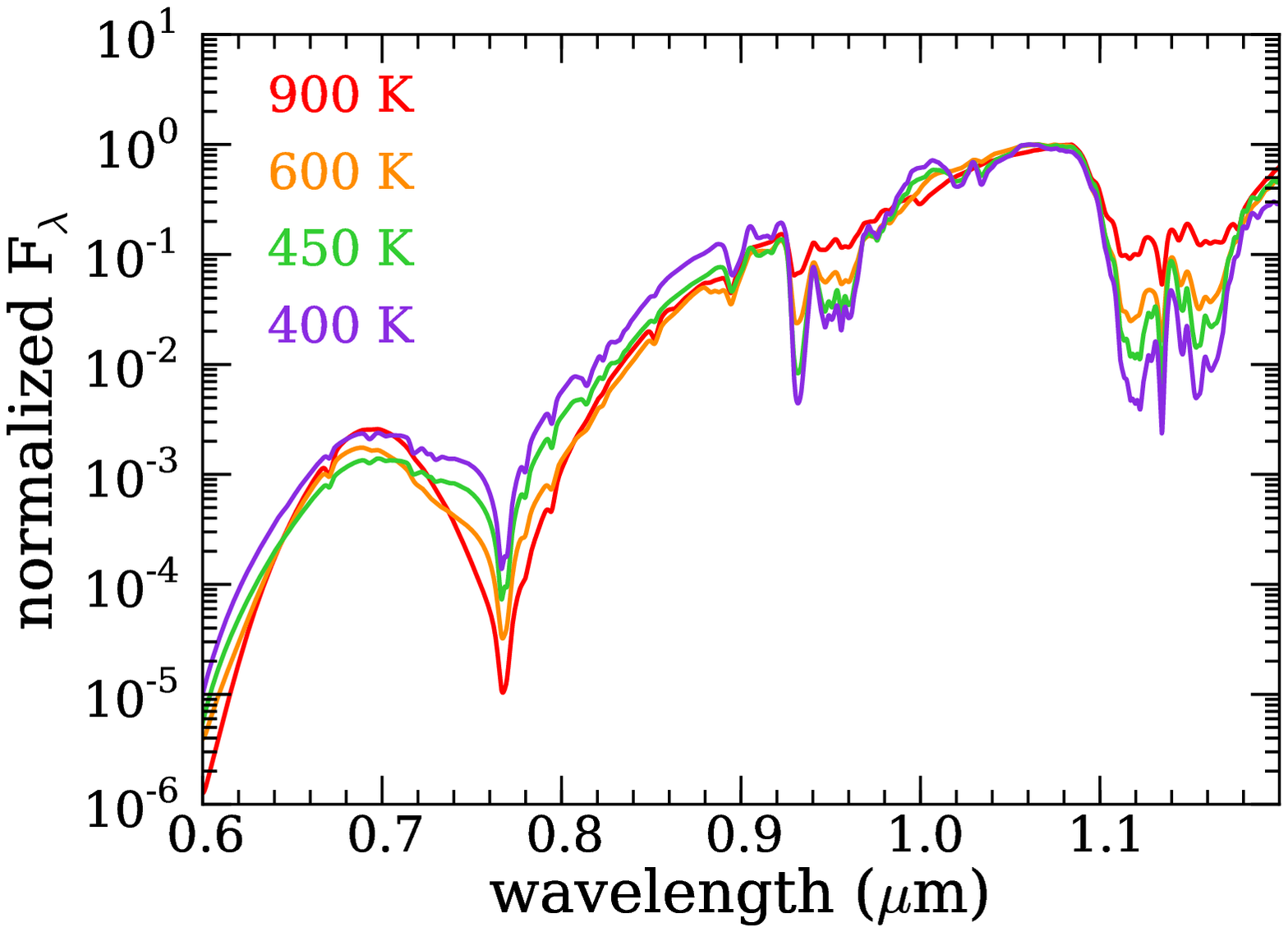}
  \caption{Shape of the red optical and \emph{Y} band over the late T to Y sequence. The spectra are normalized to the same peak flux in \emph{Y} band. The strength of the potassium feature at 0.77 \micron\ decreases as the brown dwarf cools. }
\label{alkalis}
\end{figure}

Because of the high densities in brown dwarf atmospheres, sodium and potassium bands at optical wavelengths are extremely pressure-broadened in brown dwarf spectra \cp{Tsuji99, Burrows00b, Allard05, Allard07}.  There are few other optical absorbers in brown dwarf atmospheres, so these strong pressure-broadened features shape the optical spectra of most brown dwarfs, but for Y dwarfs these atoms begin to wane in abundance. As is shown in Figure \ref{alkalis}, as sodium and potassium condense into \nas\ and KCl solid condensates, the depths and widths of the alkali features decrease. Because the slope of the optical spectrum for warmer objects is largely controlled by the pressure-broadening of these features, as they decrease in strength, the overall slope of the optical also decreases, making Y dwarfs somewhat bluer in these colors. 

\subsubsection{Gravity signatures}

Of great interest to the community in the coming years is the detection of cold planetary mass objects, either orbiting stars or free-floating \cp{Marois08, Liu13}.

\begin{figure*}[htb]
 \center    \includegraphics[width=7in]{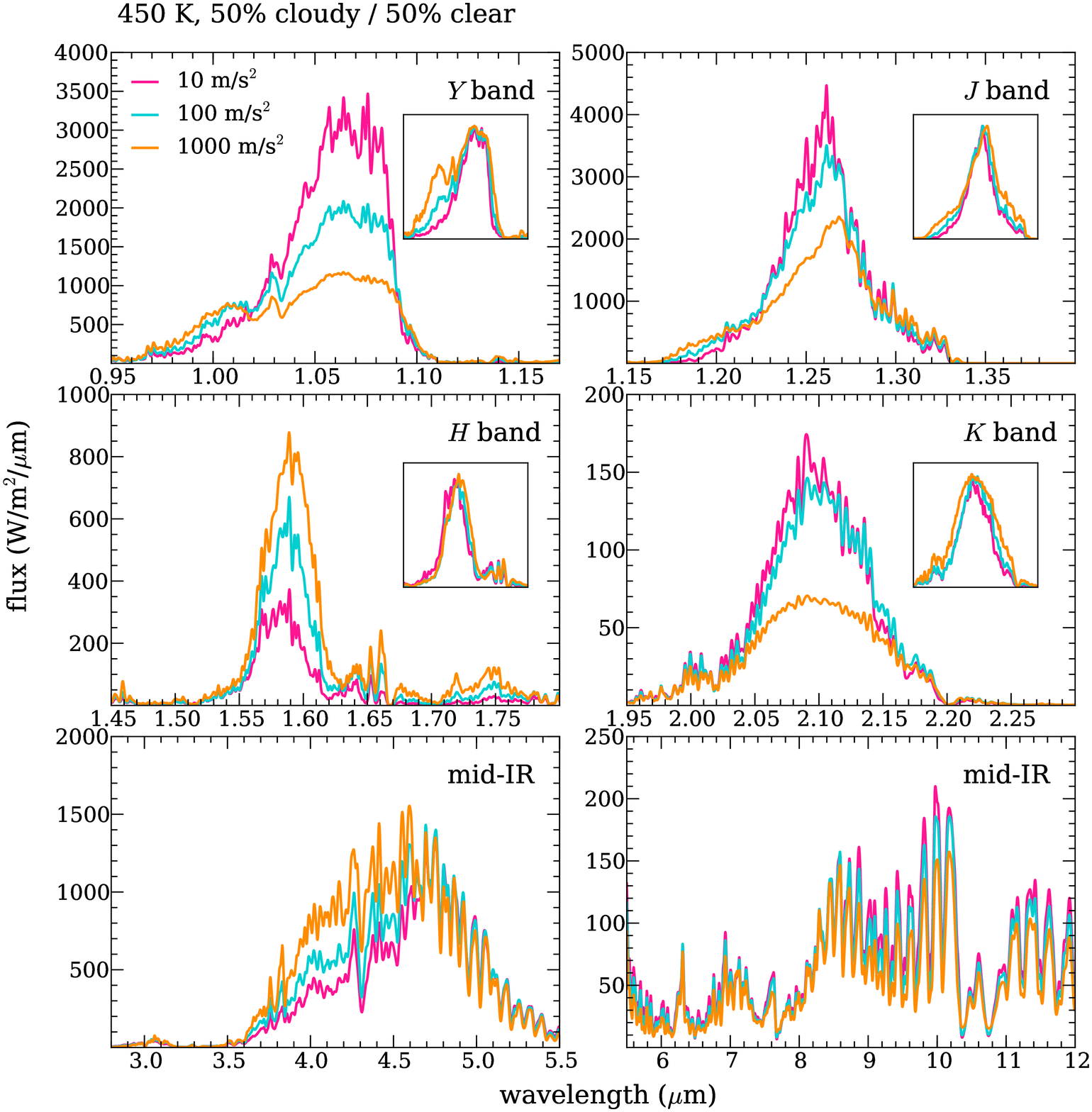}
  \caption{Gravity signatures in near- and mid-infrared. Each panel shows a wavelength range centered on a prominent molecular window, from top left, \emph{Y}, \emph{J}, \emph{H}, \emph{K}, 3--4.5 \micron, and 6--12 \micron. The inset panels for the near-IR bands show normalized version of the feature to show how the shape changes. In \emph{Y}, \emph{J}, and \emph{K}, high gravity broadens the shape of the window; between 3.5--4.6 \micron, the lower gravity spectra are more strongly influenced by absorption by PH$_3$, changing the shape of the feature. }
\label{gravsigs}
\end{figure*}

Figure \ref{gravsigs} shows potential gravity signatures for 450 K objects predicted by our models in the near-infrared ($Y$, $J$, $H$, $K$) and from 3--12 \micron. For the near-infrared bands, the inset figure shows the shapes of the bands with the peak flux in the bands normalized to the same relative height. These broad gravity signatures are largely caused by the higher pressure photospheres of higher gravity objects. 

In $Y$ and $J$ bands, the wings of the alkali bands, especially potassium, extend into these bands. These broad wings are due to the extreme pressure-broadening of the alkali lines. The higher pressures probed in the higher gravity photosphere cause less flux to emerge in both $Y$ and the blue side of $J$ band. 

As opacity sources change due to increased pressure, the temperature profile of the atmosphere adjusts; for Y dwarfs, this brightens \emph{H} band, causing more flux to emerge from that window. 

In $K$ band, the collision-induced absorption of H$_2$ is a major opacity source. This feature is quite pressure-dependent, so much like the broadened alkali bands in $Y$ and $J$ bands, the higher gravity atmosphere with the higher pressure photosphere has more total opacity in $K$ band. This decreases the amount of flux that emerges and broadens the $K$ band shape. 

The major gravity dependent feature at wavelengths longer than 3 \micron\ is between 3.5 and 4.7 \micron. At these wavelengths, the lower gravity objects have an additional absorber which changes the shape of that mid-infrared feature. This region of the spectrum is in a window between major methane and water absorption features and is the brightest peak in the near- mid-infrared spectra. PH$_3$ is a strong absorber from 4--4.6 \micron\ and somewhat gravity dependent at those wavelengths, absorbing more strongly in the lower gravity models. Additionally at these wavelengths, due to the different P--T profile at high gravity, we probe somewhat deeper, hotter layers, allowing more flux to emerge from the higher gravity model. 

\subsection{Model Photometry} \label{colors}

In order to compare to observations of the Y dwarf population, we calculate model photometry. The radii used to calculate absolute magnitudes were interpolated using  the cloud-free evolution models from \ct{Saumon08}. 

\begin{figure*}[htb]
 \center    \includegraphics[width=7in]{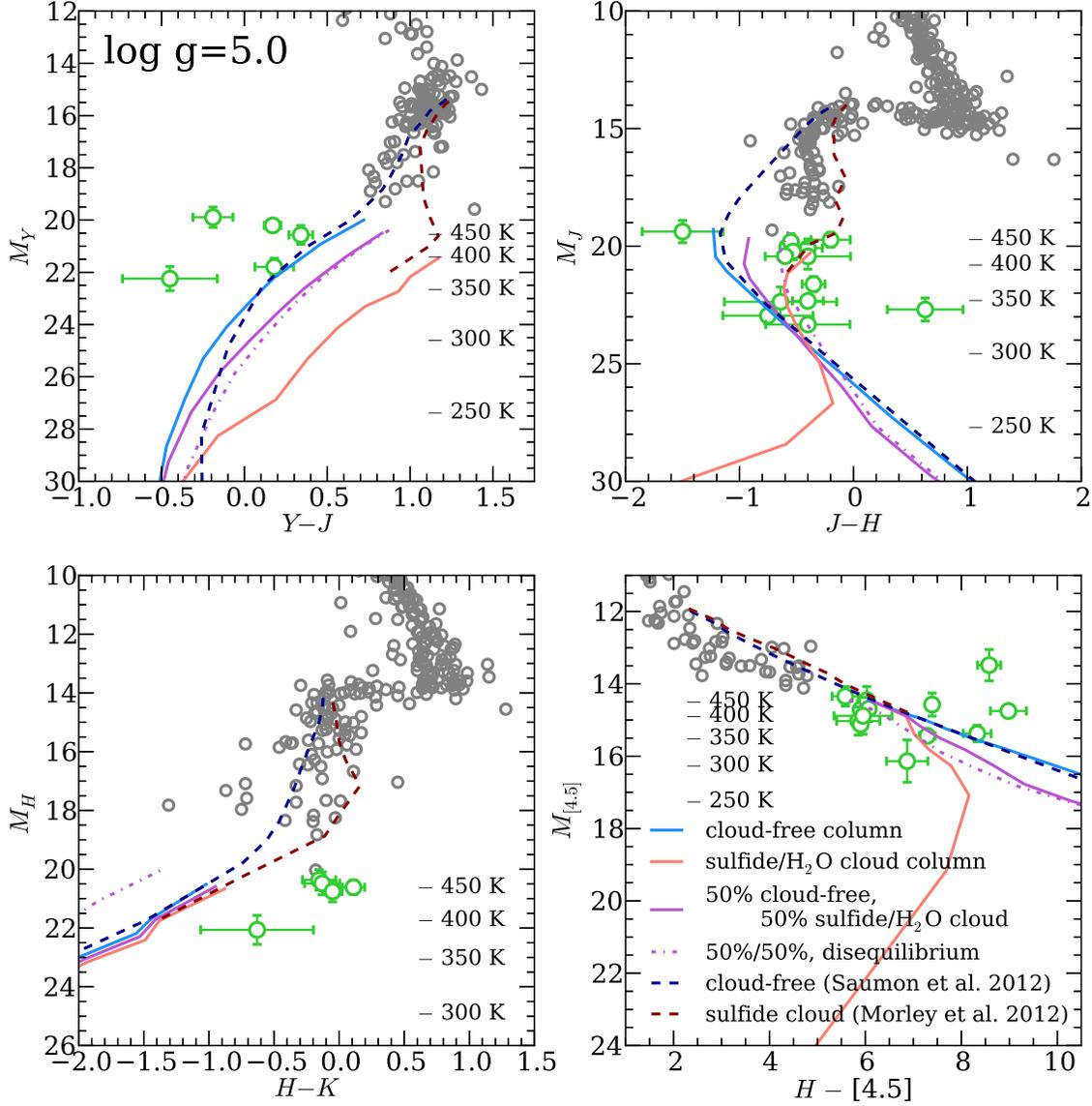}
  \caption{Color--magnitude diagrams at log g=5.0. L and T dwarfs are shown in gray, Y dwarfs are shown in green with error bars. Y dwarf parallax data is from \ct{Dupuy13, Beichman14}. L and T dwarf photometry is from \ct{Dupuy12}. The top left panel shows $Y-J$ vs. $M_Y$; the top right panel shows $J-H$ vs. $M_J$; the bottom left panel shows $H-K$ vs. $M_H$; the bottom right panel shows $H-[4.5]$ vs. $M_{[4.5]}$. The temperatures along the side show the magnitude at which the 50\% cloud-free/50\% cloudy model has that temperature (solid purple line).  }
\label{cmd-logg5}
\end{figure*}

\begin{figure*}[htb]
 \center    \includegraphics[width=7in]{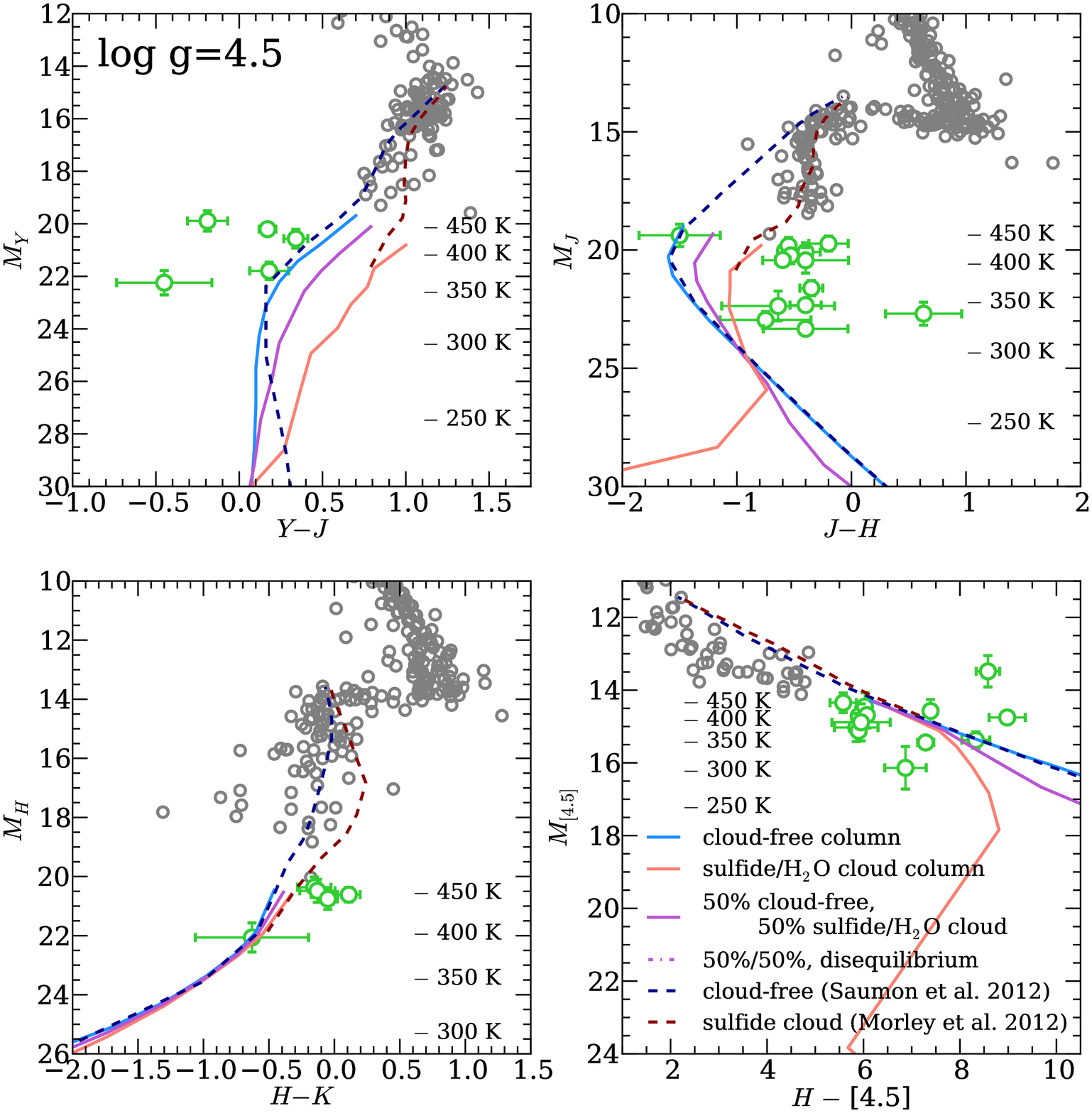}
  \caption{Color--magnitude diagrams at log g=4.5. L and T dwarfs are shown in gray, Y dwarfs are shown in green with error bars. Y dwarf parallax data is from \ct{Dupuy13, Beichman14}. L and T dwarf photometry is from \ct{Dupuy12}. The top left panel shows $Y-J$ vs. $M_Y$; the top right panel shows $J-H$ vs. $M_J$; the bottom left panel shows $H-K$ vs. $M_H$; the bottom right panel shows $H-[4.5]$ vs. $M_{[4.5]}$. The temperatures along the side show the magnitude at which the 50\% cloud-free/50\% cloudy model has that temperature (solid purple line).  }
\label{cmd-logg45}
\end{figure*}

Color--magnitude diagrams (CMDs) of the model photometry are shown for two different gravities (log g=5.0 and log g=4.5) in Figures \ref{cmd-logg5} and \ref{cmd-logg45} respectively. Each set of CMDs shows L and T dwarfs as open grey circles and Y dwarfs as green points with error bars. Models from this work and models from \ct{Saumon12} and \ct{Morley12} are shown. 

The first panel ($Y-J$ vs. M$_Y$) shows that Y dwarfs are significantly bluer in this color than the slightly warmer T dwarfs, with a 0.25 magnitude gap in color between the coolest T dwarfs and warmest Y dwarfs. This bluer $Y-J$ color is expected and indeed predicted by the models, and is due to the condensation of the alkalis into Na$_2$S and KCl clouds. As they condense out of the gas phase, the broadened alkali lines decrease in strength \cp{Marley02, Burrows03b}. Since those lines had been suppressing $Y$ band flux more than $J$ band flux, as they decrease the effect is to make $Y-J$ appear bluer. However, three of the six Y dwarfs with measured $Y$ and $J$ photometry are appreciably bluer than even the bluest models predict, and cloudy models are significantly redder. 

The second CMD shows $J-H$ vs. M$_J$. These colors are most sensitive to the sulfide and chloride clouds, which suppress the $Y$ and $J$ band flux. Since clouds tend to suppress the flux in $J$ band, the clouds make these colors redder, matching the observations of the redder Y dwarfs. The sulfide/chloride clouds wane in importance as objects cool from 450 to 325 K. For objects cooler than 325 K, the water clouds become increasingly important. Counterintuitively, the water clouds tend to make $J-H$ colors bluer; this is because of water ice is strongly scattering (and a poor absorber) in $J$ band but becomes absorbing in $H$ band (see single scattering albedo plot, Figure \ref{ssa}). Around 325 K, the cloudy color and cloud-free colors are the same. 

On the $J-H$ CMD, a line showing the effect of disequilibrium chemistry is also shown. Interestingly, this also makes the $J-H$ colors redder, mostly due to decreased absorption from NH$_3$ across the near-infrared. In reality, a combination of condensates and disequilibrium may be affecting Y dwarf spectra. 

The third panel shows $H-K$ vs. $M_K$. Somewhat puzzlingly, all the Y dwarfs except one cluster around a color of 0.0 and M$_H$ of 20.5. This behavior is quite different from the late T dwarfs, which have a large spread in $H-K$ colors. The location of this cluster is somewhat redder and brighter in $H$ band than the models predict. This could be due to a number of factors; a major contributor is the incompleteness of the methane line list used in the current study. We expect from preliminary results that the new line list from \ct{Yurchenko14} will redden these colors. 

The last panel shows the Y dwarfs where they emit the most energy: the mid-infrared. Brown dwarfs get quickly redder in $H-[4.5]$ color as they cool and the peak of the Planck function moves redward. Generally the colors of the objects seem to match the model colors relatively well, with the exception of the two reddest objects, which appear to be brighter than the models at 4.5 microns by up to 2 magnitudes.  

\section{Observing Y dwarfs with \emph{JWST}} \label{jwst}

Y dwarfs emit most of their flux in the mid-infrared. This will make their characterization from the ground extremely challenging, especially for the coldest objects. No Y dwarfs had yet been discovered during the \emph{Spitzer} Space Telescope's cryogenic mission, which ended in 2009, so the coolest brown dwarf to have a mid-infrared IRS spectrum is spectral type T7.5 \cp{Saumon06}.  \jwst\ will have unprecedented sensitivity in the near-infrared, and it will be the main tool with which we can study the coldest brown dwarfs. The two most important instruments for spectroscopic characterization will be the Near-Infrared Spectrograph (NIRSpec) and the Mid-Infrared Instrument (MIRI). 

\subsection{NIRSpec}

\begin{figure*}[htb]
 \center    \includegraphics[width=6in]{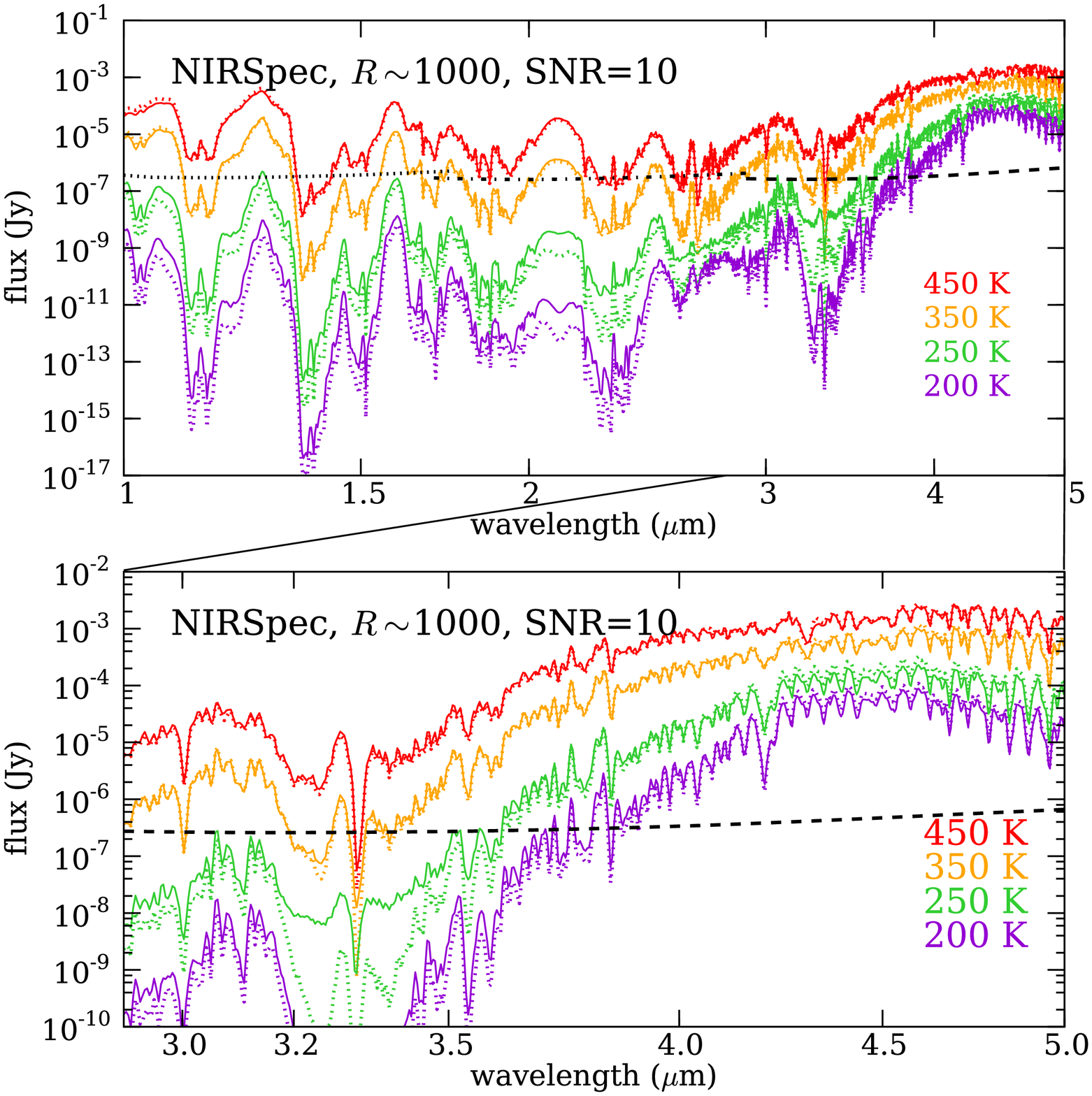}
  \caption{Model brown dwarf spectra with NIRSpec sensitivity limits. The brown dwarf spectra are scaled to represent objects 5 pc away from Earth and smoothed and binned to R$\sim$1000. All models have log g=4.5. Solid lines are the converged 50\% cloud coverage models from this work. Dotted lines are cloud-free models with the same temperature and gravity from \ct{Saumon12}. The top panel shows the sensitivity limits assuming 10$^5$ seconds of observation time in each of the three NIRSpec channels to observe a spectrum with a SNR of 10. The bottom panel zooms into the spectral region between 2.9 and 5.0 \micron. }
\label{jwst-nirspec}
\end{figure*}

NIRSpec is the most sensitive near-infrared spectrograph on \jwst\ and will be capable of moderate resolution spectroscopy (R$\sim$1000 or R$\sim$2700) from 1--5 \micron\ in 3 bands (1.0--1.8, 1.7--3.0, and 2.9--5.0 \micron\ respectively). Figure \ref{jwst-nirspec} shows the sensitivity for each of these channels; these sensitivity lines represent the faintest continuum flux observable with an integration time per channel of 10$^5$ seconds (27.8 hours) and a signal-to-noise ratio (SNR) of 10. We also show example spectra of brown dwarfs spanning a full range of Y dwarf temperatures and located 5 pc from the Earth. In the bottom panel we zoom into the band 3 spectral region, where even our coolest models (\teff=200 K) would be observable. The dotted lines show models with no clouds from \ct{Saumon12}; the solid lines show the models from this work including water and sulfide/salt clouds. 

The warmer Y dwarfs discovered to date, \teff=400--500 K, will be observable across the near-infrared bands, with the exception of the deepest absorption bands. We will be able to detect the presence or absence of strong features from water, ammonia, methane, phosphine, carbon monoxide, and carbon dioxide using this instrument, allowing us to constrain disequilibrium carbon, nitrogen, and phosphorous chemistry in Y dwarf atmospheres. 

The Y dwarfs cooler than 300 K will be too faint to observe at wavelengths from 1--3.5 \micron, as the flux from the near-infrared collapses. However, we will be able to detect these objects at high SNR between 3.5 and 5.0 \micron. The shape of this region is controlled by a variety of absorbers: ammonia and methane on the blue side, water on the red side, and possibly H$_2$S, PH$_3$ (if phosphorous chemistry is in disequilibrium), and CO and CO$_2$ (if carbon chemistry is in disequilibrium). The prominent feature at 4.2 \micron\ is an ammonia absorption feature and could be useful for determining the effective temperatures of cold brown dwarfs. This window region of the opacity will likely provide the most information about these otherwise very faint objects. 

\subsection{MIRI}

\begin{figure*}[htb]
 \center    \includegraphics[width=7in]{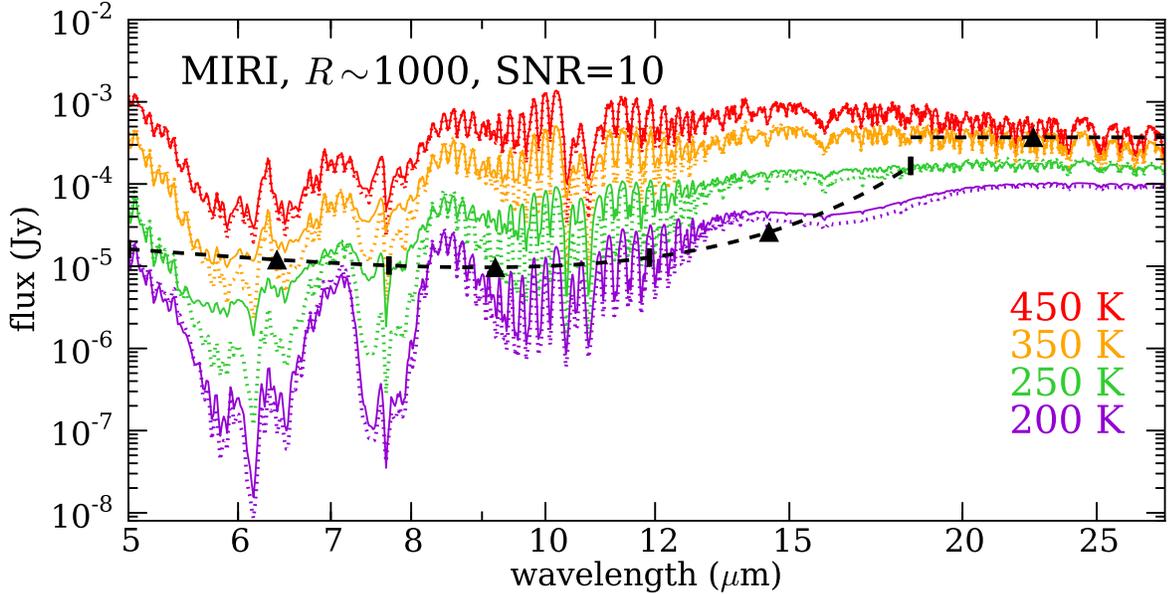}
  \caption{Model brown dwarf spectra with MIRI sensitivity limits. The brown dwarf spectra are scaled to represent objects 5 pc away from Earth and smoothed and binned to R$\sim$1000. All models have log g=4.5. Solid lines are the converged 50\% cloud coverage models from this work. Dotted lines are cloud-free models with the same temperature and gravity from \ct{Saumon12}. The sensitivity limits represent 10$^5$ seconds of observation time in each of the four MIRI channels to observe a spectrum with a SNR of 10. }
\label{jwst-miri}
\end{figure*}

MIRI will be capable of low (R$\sim$100) resolution spectroscopy from 5--14 \micron\ and moderate resolution (R$\sim$3000) spectroscopy from 5--28.3 micron. It is the only \jwst\ instrument that will observe wavelengths longer than 5 \micron\ and will be 50 times more sensitive than the \emph{Spitzer} Space Telescope. The MIRI moderate resolution spectrograph has four channels: channel 1 from 5.0--7.7 \micron, channel 2 from 7.7--11.9 \micron, channel 3 from 11.9--18.3 \micron, and channel 4 from 18.3--28.3 \micron. Figure \ref{jwst-miri} shows the sensitivity for each of these bands; like the NIRSpec sensitivity limits, these sensitivity lines represent the faintest continuum flux observable with an integration time per channel of 10$^5$ seconds (27.8 hours) and a signal-to-noise ratio (SNR) of 10, and we show example spectra of the same brown dwarf models located 5 pc from Earth. The MIRI sensitivity at 6.4 \micron\ is about 10 times less sensitive than the NIRSpec sensitivity at 5 \micron. This is due to a combination of instrumental effects, including MIRI's higher dark current, higher intrinsic spectral resolution, finer spatial sampling, lower quantum efficiency, and additional optics. To obtain the full 5--28 \micron\ spectrum without gaps, three separate observations (using three different settings of the MRS spectrograph) are needed, so the actual observing time to acquire these spectra is three times as long. 

Many of the current suite of Y dwarfs discovered to date using the \emph{WISE} data will be easily observable using this instrument. Most of these objects have temperatures between 400--500 K and are within 10 pc of the Earth \cp{Dupuy13, Beichman14}. The \teff=450 K model shown in Figure \ref{jwst-miri} is well above the sensitivity limit for 5--18.3 \micron. This will allow us to probe parts of the spectrum where the opacity is dominated by different molecules: water from 5--7.2 \micron, methane from 7.2--8.5 \micron, and ammonia from 8.5--18 \micron. Channel 4 is much less sensitive so a 450 K object is only marginally detectable from 18--28.3 \micron\ with a SNR of 10.    

The coldest brown dwarfs will push the sensitivity limits for this instrument. No objects colder than 300 K have been discovered to date, but if we find such objects, they will be quite challenging to observe. The highest SNR spectra will be from channels 2 and 3, from 7.7--18.3 \micron. This wavelength range is shaped by methane, ammonia, and H$_2$ CIA. Objects below 300 K will be only marginally detectable in channel 1 and not detectable in channel 4 at R$\sim$1000. Binning the spectra to R$\sim$300 would improve sensitivity by a factor of $\sim$3, improving the detection limit for \teff=250 K objects and still providing adequate resolution for identifying prominent molecular bands.

\subsection{NIRCam and NIRISS}

The two other main science instruments on \jwst\ are the Near-Infrared Camera (NIRCam) and Near-InfraRed Imager and Slitless Spectrograph (NIRISS). These instruments are somewhat less well-suited to the spectral characterization of the coolest brown dwarfs. NIRISS is optimized for high contrast imaging, high resolution imaging of extended sources, and transiting exoplanet measurements; it also has a wide-field R$\sim$150 slitless spectroscopy mode designed for detecting high redshift emission lines. However, for the characterization of cool brown dwarfs, sensitivity is the most important feature. 

NIRCam does have a grism mode that will be capable of 2.4--5 \micron\ R$\sim$2000 slitless spectroscopy. However, its sensitivity at those wavelengths will be somewhat lower than NIRSpec's sensitivity. The lower sensitivity is because NIRCam uses a slitless grism that is sensitive to sky background across a large field. This mode is more optimized for precision photometry and stability, making it a powerful instrument for, e.g., exoplanet transmission spectra and secondary eclipses. For the sensitivity-limited work needed to characterize Y dwarfs, NIRSpec will be a more suitable instrument. 


\section{Observing cold directly-imaged planets} \label{gpi}

Observing directly-imaged giant planets at the temperatures of Y dwarfs (200--450 K) will push the limits of current technology. Nonetheless, some current or soon forthcoming instruments specialized for high-contrast imaging will be capable of detecting such planets. If planets with masses between 1 and 10 M$_J$ are quite common, systems of a variety of ages will have planets with these temperatures. Depending on the mechanism of formation, a 10 M$_J$ planet will reach \teff=450 K at an age of 1--2 Gyr. A 5 M$_J$ planet will reach \teff=450 K in 300--600 Myr; a 1 M$_J$ planet in 20--30 Myr \cp{Fortney08b}. The current and forthcoming instruments most capable of detecting such planets are the Gemini Planet Imager (GPI), the Spectro-Polarimetric High-contrast Exoplanet Research (SPHERE), and the Large Binocular Telescope Adaptive Optics System (LBTAO). 

\begin{figure*}[htb]
 \center    \includegraphics[width=7in]{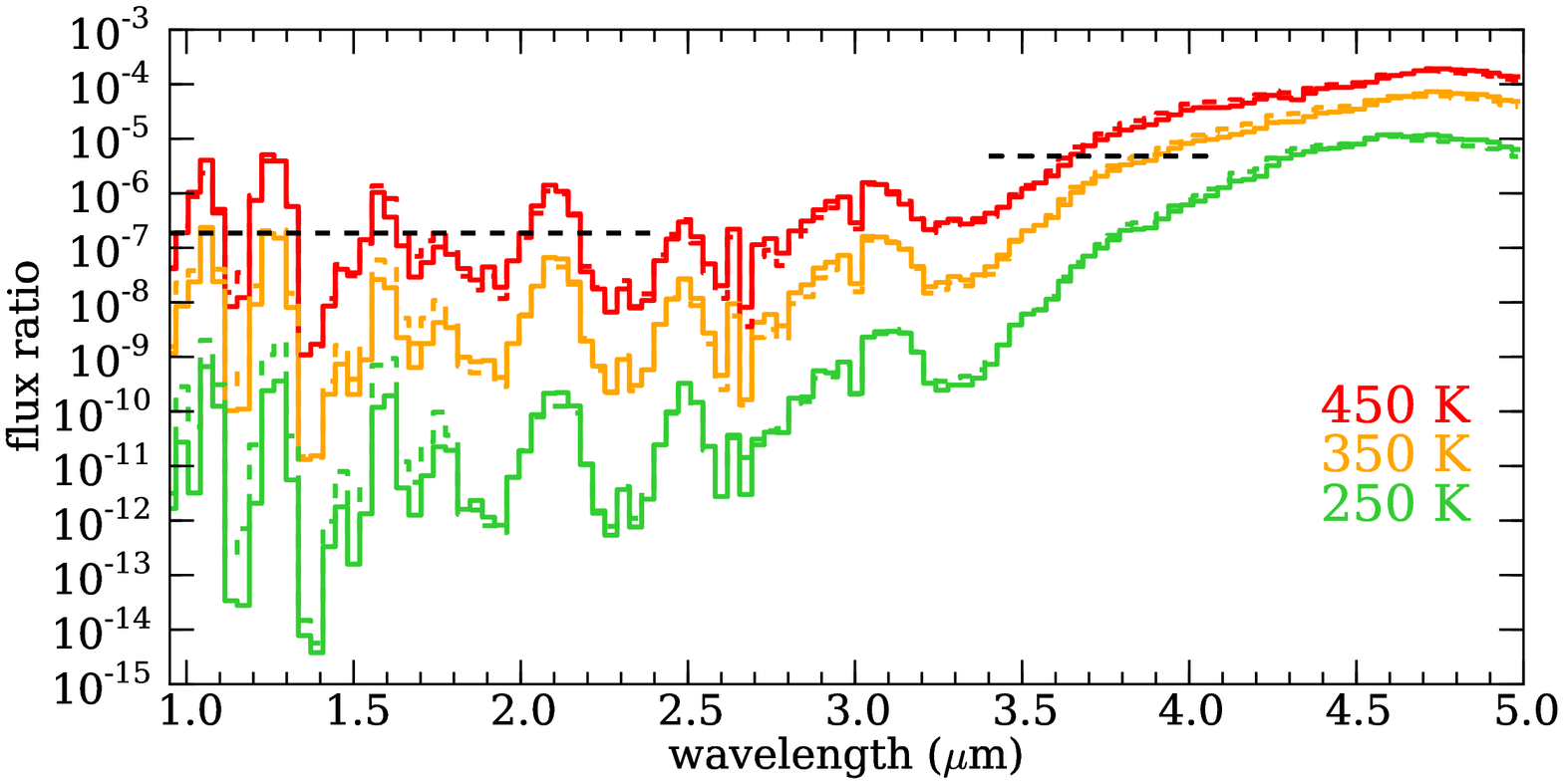}
 \caption{Spectra of model planets with \teff=450, 350, 250 K, smoothed to $R\sim$45 at 1.65 \micron. Spectra are shown as contrast ratio to a blackbody with the temperature and radius of a G0 dwarf. The black dashed lines show expected contrast around a G0 star for GPI (near-IR) and LBTAO (mid-IR) for a moderately bright star (\emph{I}=7). Solid colored lines show low gravity (log g=3.0) and dashed lines show moderate gravity (log g=4.0) for directly-imaged planets. }
\label{gpi-plot}
\end{figure*}

\vspace{20mm}

\subsection{GPI and SPHERE}

GPI \cp{Macintosh06} and SPHERE \cp{Beuzit06} are instruments designed for 8-meter class telescopes and optimized for studying young hot giant planets around bright stars. Both will have advanced adaptive optics systems and hope to achieve planet-star flux contrasts between 10$^{-6}$ and 10$^{-8}$. They will target young stars to find self-luminous planets and expect to find up to dozens of planets in their initial campaign surveys \cp{McBride11}. 

Predicted GPI contrast curves suggest that it will be capable of $5\times10^{-8}$ contrast for very bright stars ($I=5$) and 10$^{-6}$ contrast for fainter ($I=9$) targets. In Figure \ref{gpi-plot}, a representative value of 1.9$\times10^{-7}$ is shown, which is the predicted contrast at 1 arcsec separation from a 7th magnitude G0 dwarf. The models shown are binned to the resolution of GPI in \emph{H} band, $R\sim45$ at 1.65 \micron\ and are shown as the flux ratio compared to a blackbody with the temperature and radius of a G0 dwarf. The \teff=450 K planet is detectable above the contrast limits in the spectral regions where the planet is bright. Cooler planets (\teff=350, 250 K) will be too faint to observe around a G dwarf. 

\subsection{LBTAO}

The LBTAO system includes a high-contrast imaging instrument optimized for the mid-infrared. It is capable of 4.8$\times10^{-6}$ contrast in \emph{L'} band \cp{Skemer14} for a bright star and has six narrow band filters spanning 3.04 to 3.78 \micron. This spectral region is particularly useful for two reasons: first, the planet-star flux ratio is much higher for cool planets in \emph{L} and \emph{M} bands. Second, this spectral region spans the most prominent methane feature at 3.3 \micron, allowing for the characterization of a prominent atmospheric component. For example \ct{Skemer13} used this instrument to observe the HR 8799 system and find that the planets do not have strong methane absorption, inferring that methane must be in disequilibrium. 

Similar to GPI, a 450 K planet around a G dwarf would be detectable with LBTAO but a 350 or 250 K planet around a G dwarf would not be. Note that, because it operates in the mid-infrared where the sky is bright, the LBTAO system is also background limited; it can only observe objects brighter than $\sim$18th magnitude in L$'$. This limit approximately corresponds to a 350 K object at 10 pc. More distant objects will therefore be background limited; closer objects will be contrast limited. The LEECH campaign is currently using this instrument to survey stars in the solar neighborhood and discover new planets \cp{Skemer14}. 

\section{Discussion}

\subsection{Outstanding Issues}

Discrepancies between models and observations may indicate that the physics or ingredients in the model are either incorrect or incomplete. For example, there are many sources of uncertainty in the molecular opacity databases.

Alkali opacity uncertainties may affect the optical and near-infrared. The handful of Y dwarfs that have observed $Y-J$ color suggest that these objects are bluer in this color than the models. This spectral region is controlled in part by the decline of the strongly pressure-broadened alkali opacity, so this mismatch could be due to the treatment used for the alkali opacity. For these calculations, we use the line broadening treatment outlined in \ct{Burrows00b}, which is somewhat \emph{ad hoc} and potentially creates some inaccuracies in the model flux in $Y$ and $J$ bands. A calculation of the molecular potentials for potassium and sodium in these high pressure environments, as is carried out in \ct{Allard05, Allard07}, would improve the accuracy of these models. Subsolar metallicities also make $Y-J$ colors bluer \cp{Mace13a, Burrows06}, so it is important to model the opacity correctly to interpret metallicity measurements using this spectral region. 

Another known source of opacity uncertainty is in the methane line list; the list used in this study is known to be incomplete. The new line list from \ct{Yurchenko13} and \ct{Yurchenko14} has over nine billion lines and will vastly improve accuracy of the treatment of methane. 

It is possible that we may be missing important physical processes in our models that occur in actual Y dwarfs. For example, an assumption we make when calculating the spectra is that the atmospheres are in radiative--convective equilibrium. It is inevitable that real brown dwarf atmospheres have higher levels of complexity than these simple assumptions. The upper atmospheres of brown dwarfs could be heated by a similar mechanism to that which creates Jupiter's thermosphere, in which energy is deposited high in the atmosphere by dissipation of gravity waves \cp{Young97}. If indeed the upper atmosphere is hotter than radiative--convective equilibrium models, this would change the observed spectrum. Using mid-infrared spectra of L dwarfs, \ct{Sorahana14} show that several L dwarf spectra can be fit significantly better using a model that allows for this upper atmospheric heating. In fact, in our equilibrium models the [3.6]$-$[4.5] color is redder by $\sim$1 magnitude at some \teff; simple preliminary models in which we change the P--T profile of the upper atmosphere show that heating high in the atmosphere increases the flux within the methane band centered at 3.3 \micron, which makes the [3.6]$-$[4.5] color bluer, and closer to the observed colors. 

Another limitation of this study is that the models here include only solar abundances. We expect Y dwarfs and exoplanets to have a range of metallicities and potentially a range of other abundance ratios such as non-solar C/O ratios. Future work will be needed to analyze the effect of abundances on Y dwarf spectra and colors. 

We consider disequilibrium chemistry of N$_2$/NH$_3$ and CO/CH$_4$, but other molecules such as PH$_3$ and CO$_2$ may also be out of chemical equilibrium. Such unmodeled chemistry could change abundances of molecules we do include, or create additional molecules that we do not include in the calculations.

As 1D models, these calculations naturally do not include the effect of 3D dynamics on the cloud structure. The breakup of the iron and silicate clouds at the L/T transition is a source of continued study; due to dynamical processes, those clouds dissipate at higher temperatures than 1D cloud settling models predict. As the sulfide and salt clouds sink more deeply into the atmosphere as the brown dwarf cools, they may also break up and disappear from spectra at higher temperatures than our treatment predicts. Further study, including dynamical effects, will be necessary to understand this phenomenon across the brown dwarf spectral sequence.

\subsection{Comparison with Burrows et al. 2003 models} \label{burrows}

To our knowledge, the only previous comprehensive set of models analyzed and published for Y dwarfs including water clouds using a cloud model were those in \ct{Burrows03b}. Our approaches and results are overall very similar; we assume chemical equilibrium, radiative--convective equilibrium, incorporate a cloud model, and publish spectra and colors of the coolest brown dwarfs. Like \ct{Burrows03b}, we predict the growing importance of methane and ammonia absorption over the T to Y sequence, the weakening of the alkali absorption, and a reversal in the blueward trend in near-IR colors of the T dwarfs around 400 K. Our spectra differ in details due largely to changes in the line lists over the past decade; we have continuously improved our opacity database over the last ten years \cp{Freedman08}. Most relevant here, we are using updated treatments for both ammonia (which becomes very important for Y dwarfs) and H$_2$ collision-induced absorption. 

One of the conclusions presented in \ct{Burrows03b} is that the water clouds do not significantly affect the spectra of Y dwarfs and it is on this point that we differ most significantly. The differences lie in our treatment of the cloud model for the water clouds. \ct{Burrows03b} uses the cloud model presented in \ct{Cooper03}, which results in a uniform distribution of very large particles (20--150 \micron) within a single pressure scale height. Since our cloud particles are much smaller ($\sim$1-20 \micron), we have far more particles to create a cloud of the same mass. This means that when our water cloud forms, it is much more optically thick. For a more detailed comparison of the cloud models themselves see \ct{Marley03} which describes the challenges of modeling clouds in brown dwarfs and the problems inherent to different approaches.

\subsection{WISEPA J182831.08+265037.8}

WISEPA J182831.08+265037.8 (hereafter WISEP J1828+2650) is a particularly interesting object. Its near-infrared colors are inconsistent with the models and with the other Y dwarfs. The peculiarities of the near-infrared colors and comparisons to models led \ct{Cushing11} and \ct{Kirkpatrick12} to classify it as a >Y0 and >Y2 respectively; comparisons to models gave \ct{Cushing11} a temperature estimate of \teff$\le$300 K. However, for a brown dwarf that cold to have the measured mid-infrared luminosity, it would need to have an unphysically large radius, leading \ct{Leggett13} to suggest that it is an unresolved binary. 

\ct{Dupuy13} revised the parallax measurement and found that based on its luminosity, WISEP J1828+2650 likely has a temperature closer to the late T dwarfs, 500--600 K, and that even if it is a binary, those components must still be 400--500 K. 

Using the models presented here, it is not possible to fit all of the near-infrared colors of WISEP J1828+2650 using a lower gravity model. For example, the models predict that a brighter \emph{J} band flux and corresponding bluer $J-H$ color at lower gravity, due to the decreased strength of the alkali absorption in a lower gravity photosphere, whereas for this object we observe an extremely red $J-H$ color. 

Additional spectroscopic data at near- and mid-infrared wavelengths will be required to determine whether WISEP J1828+2650 is indeed a prototypical \teff$\le$300 K Y dwarf, or a peculiar version of the T8--Y0 spectral classes.

\section{Conclusions}

As brown dwarfs approach the effective temperatures of the solar system's planets, volatile clouds will form in their atmospheres. The first and most massive type of volatile cloud that forms is water ice clouds. Water ice clouds form in objects cooler than effective temperatures of $\sim$400 K. In order to converge atmospheric temperature structures self-consistently with both clouds and chemistry, we calculate models in which, like water clouds in the solar system planets, the clouds heterogeneously cover the surface (``patchy'' clouds). Our model grid covers the Y dwarf spectral class as well as giant planets with the same effective temperatures, from \teff=200--450 K and log g=3.0--5.0. 

Our main results include: 

1. While water condenses high in the atmospheres of all objects below \teff$\sim$400 K, these clouds do not become optically thick until the object has cooled to 350--375 K. This result means that for the current set of Y dwarfs warmer than 400 K, water clouds will not strongly effect their spectra. 

2. Water clouds, unlike other clouds in brown dwarf atmospheres, are very much non-gray absorbers. Using the \ct{AM01} cloud model, water ice particle sizes range from $\sim$1--20 \micron. For these particle sizes, the ice particles are strongly scattering in the optical through \emph{J} band and do not change the spectra significantly at those wavelengths. The ice particles absorb strongly in the infrared with prominent features, the strongest of which is at 2.8 \micron. 

3. H$_2$O, NH$_3$, CH$_4$, and H$_2$ CIA are the dominant opacity sources in Y dwarf atmospheres. Less abundant species such as PH$_3$ may also be observable at 4--4.6 \micron, as well as H$_2$S in \emph{H} and \emph{Y} bands and the alkalis in the optical.  

4. JWST's MIRI and NIRSpec instruments will be well-suited to characterizing cool brown dwarfs. \teff=400--500 K objects will be observable across their near- and mid-infrared spectra, and even \teff=200 K objects will be observable in the spectral window region between 3.8 and 5.0 \micron\ and at some wavelengths between 8 and 17 \micron. Existing and upcoming ground-based instruments such as GPI, SPHERE, and LBTAO will be capable of directly-imaging \teff=400--500 K planets around nearby G dwarfs.

\vspace{0mm}

\acknowledgements
The authors acknowledge Gregory Mace, Michael Liu, and the anonymous referee for comments that improved the paper. We also thank Andy Skemer for information on the observing capabilities of the LBT. We also acknowledge the Database of Ultracool Parallaxes maintained by Trent Dupuy. JJF acknowledges the support of NSF grant AST-1312545, DS acknowledges the support of NASA Astrophysics Theory grant NNH11AQ54I, and MSM and KL acknowledge the support of the NASA Astrophysics Theory and Origins Programs.

\vspace{0mm}


\end{document}